\documentclass[sigconf,nonacm]{acmart} % arxiv
\usepackage{hyperref}
\usepackage{comment}
\usepackage{type1cm} % type1 computer modern font
\usepackage{graphicx} % advanced figures
\usepackage{xspace} % fix space in macros
\usepackage{balance} % to better equalize the last page
\usepackage{booktabs} % nicer tables
\usepackage{multirow} % multi rows for tables
\usepackage[font={bf}, tableposition=top]{caption} % captions on top for tables
\usepackage{subcaption} % subfloats
\usepackage{bold-extra} % bold + {small capital, italic}
\usepackage[vlined,linesnumbered,ruled,noend]{algorithm2e} % algorithms
\usepackage{microtype} % compress text
\usepackage{siunitx} % \num for decimal grouping
\usepackage{xfrac} % nicer slanted fractions
\usepackage{mathtools} % amsmath++
\PassOptionsToPackage{hyphens}{url} % acmart compatibility
\PassOptionsToPackage{square,numbers}{natbib} % acmart compatibility
\usepackage{cleveref} % smart references
\PassOptionsToPackage{bookmarks, pdftex, colorlinks=true, pagebackref=true, backref=page}{hyperref} % acmart compatibility
\usepackage[hyperpageref]{backref} % back references
\usepackage{hyphenat} % name in single line

\usepackage{scalerel}
\usepackage{afterpage}
\usepackage{tablefootnote}
\usepackage{balance}

\usepackage[hide]{chato-notes} %hide show
\usepackage{xcolor}

\usepackage{tikz}     % For drawing on top of the image
\usepackage{lipsum}   % For placeholder text

\definecolor{cbblue}{RGB}{0, 114, 178}  % A color-blind friendly blue
\definecolor{cbred}{RGB}{213, 94, 0}    % A color-blind friendly red

\newcommand{\spara}[1]{\smallskip\noindent\textbf{#1}}

\renewcommand*\backref[1]{\ifx#1\relax \else (Cited on #1) \fi}

\AtBeginDocument{%
  \providecommand\BibTeX{{%
    \normalfont B\kern-0.5em{\scshape i\kern-0.25em b}\kern-0.8em\TeX}}}

\copyrightyear{2025}
\acmYear{2025}
\setcopyright{cc}
\setcctype{by}
\acmConference[WWW '25]{Proceedings of the ACM Web Conference 2025}{April 28-May 2, 2025}{Sydney, NSW, Australia}
\acmBooktitle{Proceedings of the ACM Web Conference 2025 (WWW '25), April 28-May 2, 2025, Sydney, NSW, Australia}
\acmDOI{10.1145/3696410.3714698}
\acmISBN{979-8-4007-1274-6/25/04}

\begin{document}

\title{Exposing Cross-Platform Coordinated Inauthentic Activity in the Run-Up to the 2024 U.S. Election}

\author{Federico Cinus}
\authornote{These authors contributed equally to this work.}
    \affiliation{
    \institution{University of Southern California, ISI, Los Angeles, USA}
    \country{}
}
\affiliation{%
  \institution{CENTAI, Turin, Italy}
  % \city{Turin}
  % \country{Italy}
  \country{}
}
\affiliation{%
  \institution{Sapienza University, DIAG, Rome, Italy}
  % \city{Rome}
  % \country{Italy}
  \country{}
}
\email{cinus@diag.uniroma1.it}

\author{Marco Minici}
\authornotemark[1]
\affiliation{
  \institution{University of Southern California, ISI, Los Angeles, USA}
  \country{}
}
\affiliation{%
  \institution{ICAR-CNR, Rende, Italy}
  % \city{Rende}
  % \country{Italy}
  \country{}
}
\affiliation{%
  \institution{University of Pisa, CS Department, Pisa, Italy}
  % \city{Pisa}
  % \country{Italy}
  \country{}
}
\email{marco.minici@icar.cnr.it}

% \authornote{Both authors contributed equally. Contact author: cinus@diag.uniroma1.it}

\author{Luca Luceri}
\affiliation{%
  \institution{University of Southern California, ISI, Los Angeles, USA}
  \country{}
}
\email{luceri@isi.edu}

\author{Emilio Ferrara}
\affiliation{%
  \institution{University of Southern California, ISI, Los Angeles, USA}
  \country{}
}
\email{emiliofe@usc.edu}

% \thanks{\textsuperscript{*}These authors contributed equally to this work}
\renewcommand{\shortauthors}{Federico Cinus, Marco Minici, Luca Luceri, and Emilio Ferrara}

\begin{abstract}
Coordinated information operations remain a persistent challenge on social media, despite platform efforts to curb them. While previous research has primarily focused on identifying these operations within individual platforms, this study shows that coordination frequently transcends platform boundaries. Leveraging newly collected data of online conversations related to the 2024 U.S. Election across $\mathbb{X}$ (formerly, Twitter), Facebook, and Telegram, we construct similarity networks to detect coordinated communities exhibiting suspicious sharing behaviors within and across platforms. Proposing an advanced coordination detection model, we reveal evidence of potential foreign interference, with Russian-affiliated media being systematically promoted across Telegram and $\mathbb{X}$. 
Our analysis also uncovers substantial intra- and cross-platform coordinated inauthentic activity, driving the spread of highly partisan, low-credibility, and conspiratorial content. 
These findings highlight the urgent need for regulatory measures that extend beyond individual platforms to effectively address the growing challenge of cross-platform coordinated influence campaigns.
\end{abstract}



\maketitle

\section{Introduction}
% !TEX root =  ./main.tex
\label{sec:intro}
Social media has evolved into a vital arena for public discourse, where individuals and communities discuss political, social, and cultural issues. 
These platforms have fostered large-scale movements like the Arab Spring and Black Lives Matter, enabling activists and citizens to amplify calls for justice and change~\cite{howard2013democracy, freelon2016beyond}.
However, social media has also been linked to polarization and radicalization, as seen in events surrounding the U.S. Capitol attack \cite{universitythe,luceri2021social}, where online platforms played a role in mobilizing and coordinating participants ~\cite{vishnuprasad2024tracking}. 
Similar cases include the Christchurch mosque shootings~\cite{fakhrurroja2019crisis} and far-right rallies across Europe~\cite{klein2019online, perez2020trend}, where social media content has exacerbated divisive sentiments and mobilized fringe communities~\cite{davey2019great}. 
The influence of online discourse has contributed to the establishment of state-sponsored information operations aimed at steering public opinion by introducing false or misleading narratives on popular platforms~\cite{nisbet2019psychology, wilson2018assembling, molter2020pandemics}. 
Online platforms actively monitor and attempt to mitigate the integrity threats posed by information operations \cite{tiktok-integrity, meta-integrity, youtube-integrity}.

Coordinated inauthentic behavior is a prominent tactic for distorting public discourse by amplifying specific viewpoints and giving the illusion of widespread support~\cite{giglietto2020detecting, khaund2021social, pacheco2021uncovering, murero2023coordinated, cima2024coordinated}. 
These activities typically involve synchronized posting and co-activity behavior patterns~\cite{oates2019kremlin, pacheco2021uncovering, luceri2024unmasking}--- such as similar retweets, hashtag sequences, link sharing--- designed to push particular narratives to the forefront of public attention.
Studies have also documented how these coordinated behaviors contribute to detrimental social effects, such as the spread of propaganda~\cite{hristakieva2022spread}, conspiracy theories~\cite{tardelli2024multifaceted}, and the promotion of disinformation~\cite{weber2021amplifying}. Moreover, coordinated online actions have been linked to heightened toxicity in online conversations~\cite{loru2024influence} and the dissemination of extremist ideologies~\cite{kursuncu2021bad, vishnuprasad2024tracking}.

In recent years, research has increasingly examined social media’s role in political elections. 
Studies on Brazilian elections highlight how bots and coordinated networks shape public opinion, raising concerns on computational propaganda~\cite{woolley2017computational, pacheco2024bots}. 
Similarly, research on European elections has shown how misinformation campaigns spread across social media can alter public perceptions, as seen in countries such as France, Germany, and Italy~\cite{ferrara2017disinformation, lovari2020spreading, baqir2024news,nogara2024misinformation}. 
In the United States, investigations reveal how coordinated bot activity and foreign influence operations distorted online discussions and amplified divisive content across election cycles~\cite{bessi2016social, howard2018algorithms, luceri2019evolution, tardelli2024multifaceted,luceri2021down,ferrara2020characterizing}. 
However, most of this literature focuses on identifying such operations within individual platforms, predominantly Twitter, with limited investigation across the broader information space \cite{gatta2023interconnected}.

\subsection*{Contribution of this work}
In this work, we expand on current research in coordination detection by broadening our scope to include the identification of coordinated inauthentic activity (CoIA) spanning multiple platforms. 
We adhere to the definition of ``coordinated and inauthentic behavior'' provided by Meta\footnote{about.fb.com/news/tag/coordinated-inauthentic-behavior/} and Twitter/$\mathbb{X}$\footnote{help.x.com/en/rules-and-policies/authenticity}.
We build on, adapt, and advance state-of-the-art coordination detection techniques to identify intra- and cross-platform CoIA that promotes external web domains and amplify specific narratives in the context of the 2024 U.S. Presidential Election. We leverage a large-scale dataset covering election-related online conversations spanning several platforms \cite{blas2024unearthing,memo6}, including Twitter/$\mathbb{X}$, Facebook, and Telegram, collected during the run-up months to the Presidential Election. Combining insights from computational techniques for CoIA detection and language models for content analysis, we aim to answer the following Research Questions (RQs):

\begin{itemize}
\item[\textbf{RQ$_{1}$:}] \textbf{Web Domain Promotion.} 
\textit{Do we observe intra- and cross-platform CoIAs aimed at redirecting traffic towards specific web domains? %How do these accounts coordinate within and across different platforms? 
        What are the characteristics of coordinated accounts, and which specific domains do they promote?}

\item[\textbf{RQ$_{2}$:}] \textbf{Content Amplification.} 
 \textit{Do we observe CoIAs pushing specific narratives? What are the characteristics of coordinated accounts, and which specific topics do they amplify?}
        % they push controversial and conspiracy topics
\item[\textbf{RQ$_{3}$:}] \textbf{Engagement \& Impact.} 
\textit{What level of engagement do CoIAs generate across various platforms? And how does it compare to the engagement garnered by organic users?}
\end{itemize}

Leveraging our multi-platform dataset and advancing CoIA detection techniques, we uncover multiple networks of coordinated accounts performing CoIA.
We analyzed the textual content and link-sharing behaviors of these accounts, identifying the narratives they aim to amplify and the external domains they direct traffic to, assessing both their credibility and ties to foreign and domestic entities. Our findings reveal coordinated efforts to promote Russian-affiliated media across Telegram and $\mathbb{X}$, with highly partisan, low-credibility content systematically amplified by these networks. Conspiracy theories surrounding public health, the environment, and political topics, such as immigration and geopolitical tensions, were prominently featured. Notably, QAnon-related narratives were especially prevalent on Telegram, with coordinated accounts driving much of the discussion. Furthermore, we assessed the prevalence of AI-generated content produced by coordinated actors and the level of engagement their content attracted. We found that coordinated actors on Telegram relied on AI-generated content significantly more than organic users, while the opposite trend was observed on Facebook. This study sheds new light on cross-platform coordination efforts related to the upcoming U.S. Presidential Election, revealing the complex dynamics of influence operations. These findings underscore the urgent need for regulatory measures that go beyond individual platforms to effectively tackle the growing challenge of cross-platform coordinated inauthentic activity.
% \clearpage

\section{Related Work}
% !TEX root =  ./main.tex
\label{sec:related}

\noindent{\textbf{Influence Campaigns Interfering with U.S. Elections.}}
A growing body of research has examined the influence of coordinated disinformation campaigns in U.S. elections, particularly efforts by foreign and domestic actors to manipulate social media during the 2016 and 2020 Presidential Elections~\cite{sharma2022characterizing,ferrara2020characterizing,bessi2016social,vishnuprasad2024tracking}.
For example, the Russian Internet Research Agency (IRA) was linked to a large-scale disinformation campaign during the 2016 election, deploying thousands of social bots and state-sponsored trolls to promote divisive narratives and foster discord \cite{badawy2019characterizing,badawy2018analyzing}. The activity of these inauthentic actors has been extensively studied on Twitter \cite{zannettou2020characterizing,luceri2020detecting,addawood2019linguistic,luceri2019red}, with fewer studies examining their presence on other platforms \cite{zannettou2019let,zannettou2019disinformation}.

Similarly, the 2020 U.S. Presidential Election witnessed the propagation of false claims and conspiracy theories, including narratives about voter fraud and COVID-19 misinformation \cite{ferrara2020characterizing}. Examples include the ``Stop the Steal'' movement, which spread across platforms like Twitter and Facebook, inciting allegations that the election results were fraudulent \cite{universitythe,vishnuprasad2024tracking}. Coordinated activity surrounding this narrative was widespread, often involving amplification techniques such as automated retweets or sharing similar content across accounts to create an illusion of widespread consensus \cite{luceri2021social}.

\noindent{\textbf{Detecting Coordinated Inauthentic Activity.}}
Coordinated inauthentic activity has been identified as a predominant tactic in spreading disinformation and conspiratorial content \cite{luceri2024unmasking,Pacheco_2020}. 
While not always malicious, coordination can also occur in legitimate organizing efforts, such as social movements. However, in disinformation campaigns, it is typically characterized by manipulative actions aimed at amplifying highly partisan content~\cite{pacheco2021uncovering,nizzoli2021coordinated}.

The detection of such deceptive, orchestrated efforts has evolved to address both automated and human-coordinated activities through advanced machine learning and network-based approaches. Traditionally, machine learning focused on identifying automated activities, such as bot-like behavior, and distinguishing them from human actions \cite{Yang_2019, cresci2016dna}. Recently, attention has shifted toward characterizing suspicious human-operated accounts, with research emphasizing content-, behavioral-, and sequence-based methods to detect coordinated actions by state-sponsored trolls \cite{ Hristakieva_2022, alizadeh2020content, luceri2024leveraging, luceri2020detecting, nwala2023language, ezzeddine2022characterizing}.

Network-based detection
% , however, 
has gained prominence due to its ability to reveal coordinated behavior by constructing networks that highlight similarities in user actions, such as shared content, hashtags, or synchronized posting times \cite{Pacheco_2020, pacheco2021uncovering, nizzoli2021coordinated, mannocci2024detection, magelinski2022synchronized, luceri2024unmasking}. Typical network-based metrics, such as centrality measures or edge weights, help identify clusters of users involved in coordinated Information Operations (IOs) within individual platforms, mainly Twitter \cite{luceri2024unmasking, vishnuprasad2024tracking, tardelli2024temporal}.
% These networks are analyzed using properties like node centrality and edge weight, which help identify clusters of users involved in coordinated Information Operations (IOs) \cite{luceri2024unmasking, vishnuprasad2024tracking, tardelli2024temporal}. 
% This approach uncovers automated and human-coordinated activities, offering key insights into influence campaign tactics and structure.

While previous work analyzed cross-platform harmful dynamics~\cite{ng2022cross,gatta2023interconnected,zannettou2019disinformation}, our approach introduces a fine-grained unsupervised coordination detection framework by combining multiple network measures (edge weights and node centralities) to identify highly coordinated accounts engaged in CoIA \textit{across} multiple platforms.
% Unlike existing methods, which have explored dynamics between mainstream and fringe platforms, 
% Further, our work provides a characterization of users driving CoIA across a broader range of mainstream social media platforms.
% our work also leverages state-of-the-art language models and AI-generated content detectors to provide a nuanced characterization of users across a broader range of mainstream social media platforms.
% \clearpage

\section{Data}
% !TEX root =  ./main.tex
\label{sec:data}
We collected data to analyze online discourse surrounding the 2024 U.S. Presidential Election across multiple social media platforms. 
Our data collection spans May and June 2024 and covers three major online platforms: Facebook, $\mathbb{X}$ (formerly Twitter), and Telegram. 
Each platform was queried for posts or messages containing specific election-related keywords (see Appendix~\ref{app:dataset}).

Facebook data was collected through (the now defunct) Crowdtangle, which offered access to public posts of groups and pages. 
For $\mathbb{X}$, we gather publicly available information, including original tweets, retweets, replies, and quotes, retrieved via the platform’s web interface. 
Finally, Telegram data was gathered using the Telegram API, allowing the extraction of public chats’ details, metadata, messages, and message attachments. Details on the $\mathbb{X}$ and Telegram data collection infrastructure can be found in \cite{blas2024unearthing,balasubramanian2024public}.

To ensure consistency across all three platforms, we applied an identical filtering criterion, restricting the dataset to posts and messages published between May 1, 2024, and June 30, 2024.

Table~\ref{table:dataset_overview} presents a summary of the datasets, including the number of accounts,\footnote{We will use the term \textit{accounts} to also refer to Facebook pages, Facebook groups, and Telegram channels.} posts, and unique URLs for each platform.

\begin{table}[ht]
\centering
\renewcommand{\arraystretch}{1.2}
\begin{tabular}{lrrrr}
\toprule
\textbf{Platform} & \textbf{\shortstack{Pages \\ Accounts \\ Channels}} & \textbf{\shortstack{Posts \\ Tweets \\ Messages}} & \textbf{URLs} & \textbf{Domains} \\
\midrule
Facebook & 6,137 & 46,310 & 15,009 & 5,247 \\
$\mathbb{X}$/Twitter & 178,379 & 6,021,428 & 582,052 & 35,922 \\
Telegram & 15,537 & 4,309,880 & 2,087,078 & 183,924 \\
\bottomrule
\end{tabular}
\caption{Dataset statistics across platforms.}
\label{table:dataset_overview}
\end{table}
% \clearpage

\section{Methodology}
% !TEX root =  ./main.tex
\label{sec:methods}

\subsection{Exposing CoIA within and across Social Media Platforms}
Coordinated accounts employ various strategies to run campaigns; here, we analyze two key approaches: promoting specific URLs (or web domains) and generating highly similar textual content to amplify topics. The former creates the illusion of public consensus by artificially boosting links to external webpages, mock websites, and other social media networks~\cite{minici2024uncovering,gabriel2023inductive}. 
The latter manipulates platform feed algorithms by pushing specific keywords or hashtags, attempting to boost trending topics and inflate content popularity~\cite{pacheco2021uncovering,vishnuprasad2024tracking}.
We refer to these orchestrated efforts as \textit{web domain promotion} and \textit{content amplification}, respectively. 

Since URL-sharing follows a consistent mechanism across platforms, we analyze\textit{ web domain promotion} both within and across platforms, using unique URLs embedded in posts to track coordination. In contrast, textual similarity is analyzed only within individual platforms due to significant variability in content formats and constraints. Structural differences, such as character limits and messaging styles (e.g., Telegram chats vs. Facebook and $\mathbb{X}$ posts), make cross-platform textual analysis unreliable.

\subsubsection{Detection of Web Domain Promotion}
\label{sec:url_campaigns_method}
To detect orchestrated campaigns promoting specific web domains, we analyze user similarities based on shared URLs, including both platform-specific and outbound links.
To ensure data quality and reduce noise, we applied several preprocessing steps.
First, we set a minimum activity threshold, requiring each user to share at least 10 unique URLs. This criterion, consistent with prior work~\cite{luceri2024unmasking,pacheco2021uncovering}, filters out low-engagement users who could alter the detection of coordinated accounts.
Next, we expanded shortened or obfuscated URLs using a URL expansion library\footnote{\href{github.com/dfreelon/unspooler}{github.com/dfreelon/unspooler}}, a common practice in the literature~\cite{lasser2022social}.

\smallskip

\noindent{\textbf{Co-URL similarity network.}} 
We constructed a bipartite user-URL network, where users connect to URLs they share. In the network’s adjacency matrix, rows correspond to users and columns to URLs.
Following prior work~\cite{luceri2024unmasking,pacheco2021uncovering}, we applied Term Frequency-Inverse Document Frequency (TF-IDF) to represent the user-URL matrix. To prevent bias toward overly frequent URLs, we set a maximum ``document'' frequency at the 90th percentile and a minimum threshold of five occurrences per URL. This filters out both extremely rare and overly common URLs, ensuring a more meaningful set of URLs.

The bipartite user-URL network is transformed into a user-to-user similarity network by comparing pairwise co-shared URLs. Specifically, this similarity network is constructed by computing the pairwise cosine similarity between user TF-IDF vectors within the bipartite graph. This measure serves as the edge weight, quantifying the degree of similarity between users based on their shared URLs.

\smallskip

\noindent{\textbf{Intra-platform and cross-platform detection.}} 
Being agnostic to the platform under scrutiny, URLs serve as common entities that can be analyzed across different platforms to identify potential cross-platform campaigns.
Therefore, we constructed two types of co-URL networks: intra-platform and cross-platform. For the intra-platform network, we computed cosine similarity between all pairs of users within the same platform. For the cross-platform network, we calculated pairwise similarity only between users from different platforms, linking users based on their shared URL patterns.

\smallskip

\noindent{\textbf{Extraction of network properties for coordinated accounts detection.}} To identify online coordination, we build upon and integrate state-of-the-art strategies that detect CoIAs by filtering either low-weight edges or peripheral nodes from the similarity network. 
The first approach \cite{pacheco2021uncovering} emphasizes similarity strength, filtering out low-weight edges that likely represent spurious similarities.
The second approach \cite{luceri2024unmasking} focuses on the breadth of similarities, pruning nodes based on their centrality, i.e., the number of connections in a similarity network. This approach assumes that coordinated activity involves multiple accounts working together, which in a similarity network is expected to be reflected as a prominent \textit{collective similarity}, with coordinated accounts linked to many well-connected nodes. As demonstrated by \citet{luceri2024unmasking}, this makes eigenvector centrality an effective metric for detecting coordinated users.

Here, we extend this notion of collective similarity by considering the density of the similarity network.
Network density quantifies how close a network is to full connectivity and, in this scenario, serves as a precise measure of collective similarity. 
Network density helps us combine node and edge filtering, balancing the breadth and strength of similarities for a more comprehensive coordination detection approach. This represents a conservative approach, which is especially important since these techniques have not yet been applied to platforms like Telegram and Facebook, where coordination dynamics may differ significantly.

\smallskip

\noindent{\textbf{Density-based unsupervised network dismantling.}} 
Our innovative approach leverages network density as a key variable to modulate the combination of node and edge filtering techniques. Specifically, edge weight and node eigenvector centrality are used jointly to identify groups of coordinated accounts by filtering out edges and nodes that fall below similarity and centrality thresholds, respectively. To achieve this in a fully unsupervised manner, we introduce a grid search method that systematically explores the two-dimensional parameter space of node centrality and edge similarity quantiles. The grid search evaluates network density under various threshold combinations, progressively filtering nodes and edges while measuring their impact on the density of connected components within the filtered similarity graph.

We hypothesize that as the filtering thresholds increase, a transitional phase in component density will emerge, signaling the presence of potentially coordinated accounts. To ensure robustness, we adopt a conservative approach, focusing on the minimum density among all connected components in the filtered similarity graph. By prioritizing the component with the lowest density, we ensure that all remaining components exhibit even higher densities, minimizing false positives—i.e., reducing the risk of misclassifying legitimate users as coordinated actors. Filtering thresholds are then selected based on the transitional phase of the lowest-density component, enabling the identification of highly suspicious CoIA. Details on the parameter selection are provided in the Appendix.  

\subsubsection{Detection of Content Amplification}
\label{sec:content-amplification}
Coordinated actors may employ a variety of tactics to achieve their goals. A common tactic is to artificially amplify content on specific topics to create the appearance of widespread grassroots support and manipulate platforms' feed algorithms~\cite{mannocci2024detection,vishnuprasad2024tracking,Pacheco_2020}. To uncover content amplification, we construct a Text Similarity Network (TSN), where nodes represent users linked by the similarity of the content they share. This TSN is then employed to identify a subset of users exhibiting suspiciously high similarity, potentially indicating coordinated behavior \cite{pacheco2021uncovering}.

The initial step in constructing the TSN involves preprocessing the raw data to ensure the results are meaningful. 
In line with standard practices~\cite{luceri2024unmasking}, we exclude retweets and remove punctuation, stopwords, emojis, URLs, as well as any content with fewer than four words. 
To capture the semantic nuances of the content, we embed all text data using the SentenceTransformer model \textit{stsb-xlm-r-multilingual}\footnote{\href{huggingface.co/sentence-transformers/stsb-xlm-r-multilingual}{huggingface.co/sentence-transformers/stsb-xlm-r-multilingual}}, and then calculate the average cosine similarity between pairs of users.
To avoid spurious correlations, we consider text similarity valid only if it occurs within a one-day sliding window. For each window, we calculate the text similarity between the shared content of user pairs, setting the edge weight in the TSN to the average similarity observed across all windows.

To identify coordinated accounts, we apply established thresholds from the literature~\cite{pacheco2021uncovering, luceri2024unmasking}. Specifically, we filter out edges with similarity below 0.95 and classify users as coordinated if their nodes rank in the top 0.5\% by eigenvector centrality.
Due to substantial variability in textual content sharing, driven by platform-specific constraints, such as character limits on some platforms, and differences in the nature of messaging (e.g., Telegram chats versus Facebook and $\mathbb{X}$ posts), our
analysis of content amplification is focused only on intra-platform CoIA.

\subsection{Characterizing Content Pushed by CoIA}
In this section, we outline the methods used to characterize textual content amplified by coordinated actors. This characterization encompasses topic analysis, AI-generated content detection, and an assessment of content credibility.

\subsubsection{Topic Analysis}
\label{sec:topic-analysis}
Coordinated accounts often amplify specific narratives or themes to steer public attention toward polarized, inflammatory, or misleading discussions. To uncover the agendas these accounts seek to promote, we use BERTopic~\cite{grootendorst2022bertopic}, a state-of-the-art tool for topic extraction. 
% For a detailed description of BERTopic’s methodology, we refer readers to~\cite{grootendorst2022bertopic}. 
This approach also helps identify patterns in shared content, improving our understanding of coordinated activities and their intended impact on public discourse.
Once coordinated accounts are identified using the methodology outlined in \S\ref{sec:content-amplification}, we apply BERTopic to the entire content corpus. This allows us to map the topics promoted by coordinated accounts and compare them with those shared by organic users (i.e., users not classified as coordinated). We choose BERTopic over alternatives such as LDA or GPT-based approaches because it offers an effective balance between accuracy and scalability.

\subsubsection{AI-Generated Content Detection} 
AIGC has been linked to nefarious activities \cite{ferrara2024genai}, such as orchestrating malicious campaigns.
To identify AI-generated text, we use the approach introduced by~\cite{dmonte2024classifying}, which is designed to identify AI-generated content on $\mathbb{X}$. We initially trained a RoBERTa\footnote{\url{huggingface.co/FacebookAI/roberta-base}} model on their $\mathbb{X}$ dataset and then evaluated its performance on datasets from other platforms by developing a new validation set containing approximately 2,000 samples.
Results showed precision values ranging from 0.87 to 0.97 in detecting AI-generated content within the validation set, aligning with our goal of prioritizing precision over recall to minimize false positives (i.e., misclassified legitimate users). A detailed description of how we built the validation set and the full results are available in the Appendix.

\subsubsection{Credibility Assessment}
We assess the credibility of web domains using Media Bias/Fact Check (MBFC)---an independent watchdog that rates news outlets on a 6-point factuality scale, ranging from Very Low to Very High. 
For each post in our datasets, we systematically extract, expand, and parse all embedded URLs, checking whether the URL belongs to a low- or high-credibility domain.

Given recent news\footnote{www.bbc.com/news/articles/c8rx28v1vpro}\footnote{www.state.gov/u-s-department-of-state-takes-actions-to-counter-russian-influence-and-interference-in-u-s-elections/} and prior instances of interference by Russian agencies in U.S. elections \cite{badawy2019characterizing}, we also examined the potential amplification of Russian media outlets. Following a similar approach to previous work \cite{pierri2023propaganda}, we utilize the VoynaSlov dataset \cite{park2022challenges} to obtain a list of 23 state-affiliated Russian websites. We then check whether the extracted URLs link to one of these web domains.

Finally, we employ a similar approach to identify content linked to conspiracies, such as QAnon, given the prevalence of fringe and conspiratorial theories during the 2020 US Election, and its aftermath \cite{sharma2022characterizing,vishnuprasad2024tracking}. Building on the methodology of \cite{wang2023identifying}, we detect posts sharing QAnon content by utilizing a list of keywords commonly associated with the conspiracy, as outlined by \cite{sharma2022characterizing}. 
% \clearpage

\section{Results}
% !TEX root =  ./main.tex
\label{sec:results}

This section presents the results obtained from the detection of online coordination\footnote{Code publicly available at \url{github.com/mminici/USElection-CoIA-TheWebConf25}}. 
We first present the results for the detection of \textit{web domain promotion} and \textit{content amplification}. Following the identification of these campaigns, we examine the content pushed by coordinated networks and the engagement received by suspicious actors focusing on several interaction and engagement metrics.

\subsection{Detecting CoIA for Web Domain Promotion (RQ1)}

We detect CoIAs that push web domains within and across platforms as described in Section \ref{sec:url_campaigns_method}. 
We begin by presenting results for each coordinated network within individual platforms. Following this, we analyze the inter-platform CoIA.

\begin{figure}[t!]
    \centering
    \includegraphics[width=1.\columnwidth]{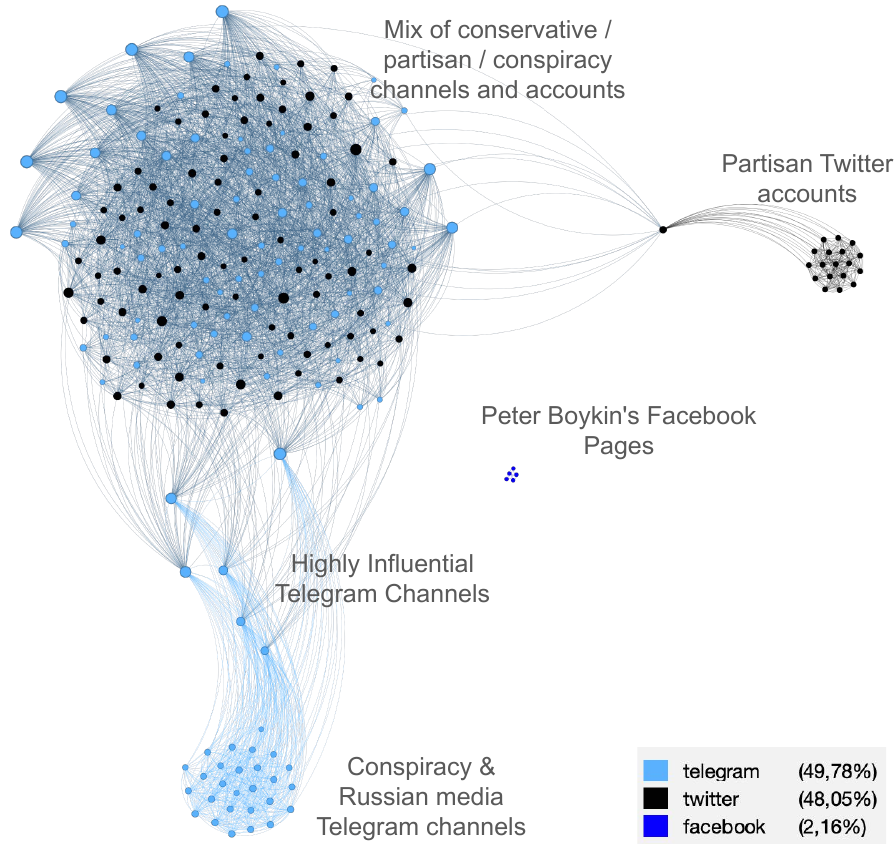}
    \caption{Cross-platform coordination network illustrating user coordination across the three social media platforms. The percentages at the bottom right indicate the proportion of users from each platform within this network.}
    \label{fig:cross-network}
\end{figure}

To provide a comprehensive view, we created a \textit{cross-platform network} by combining the intra- and inter-platform coordinated networks into a unified network. In this network, nodes are connected if they are linked in either the intra- or inter-platform coordinated networks. The primary result of this analysis is displayed in Figure~\ref{fig:cross-network}, illustrating coordinated activity both within and across the three platforms under study.

\subsubsection{Intra-Platform Coordination.} We first analyze each intra-platform CoIA based on the co-URL similarity networks extracted from each platform separately.

\smallskip

\textbf{Telegram:} We identified 33 highly coordinated channels co-sharing URLs to web domains with a partisan slant and low factuality. As shown in the Table~\ref{tab:telegram-domain-shares}, the most frequently shared domains are predominantly right-leaning and of low credibility, as assessed by MBFC\footnote{\url{mediabiasfactcheck.com/}} and consistent with previous findings~\cite{biswas2024political}. 
Notably, the table highlights the presence of a Russian state-controlled media outlet (RT.com) and its video-on-demand subsidiary (Ruptly.tv), which could indicate a foreign effort to interfere in the election. 
Both RT.com and Ruptly.tv have previously been accused of orchestrating campaigns to influence U.S. Elections via social media\footnote{www.nbcnews.com/politics/2020-election/facebook-blocks-russia-backed-accounts-other-sites-keep-churning-out-n1242683}.

Additionally, this CoIA promotes a far-right website (thegatewaypundit.com) and an extremist-friendly video platform (odysee.com), suggesting ties to fringe ideas. 
A manual review of these Telegram channels, including their profile descriptions (see Table~\ref{tab:telegram-bios} in the Appendix) and shared messages (see Table~\ref{tab:5-messages-telegram} in the Appendix), reveals a strong prevalence of content and accounts promoting conspiracy theories, particularly those related to COVID vaccines, 5G, and alternative news media.

\begin{table}[t]
    \footnotesize
    \centering
    \setlength{\tabcolsep}{8pt}
    \renewcommand{\arraystretch}{1.2}
    \begin{tabular}{c|c|c|c}
    \toprule
    \textbf{Domain}  & \textbf{Shares} & \textbf{Factuality} & \textbf{Leaning} \\
    \midrule
    \href{www.ruptly.tv}{ruptly.tv} & 2,117 & Mixed & RIGHT-CENTER \\
    \href{www.rt.com}{rt.com} & 1,941 & Very Low & RIGHT-CENTER \\
    \href{odysee.com}{odysee.com} & 1,602 & Low & RIGHT CONSPIRACY \\
    \href{www.dailymail.co.uk}{dailymail.co.uk} & 1,159 & Low & RIGHT \\
    \href{www.thegatewaypundit.com}{thegatewaypundit.com} & 746 & Very Low & EXTREME RIGHT \\
    \bottomrule
    \end{tabular}
    \caption{Top-5 domains shared in Telegram by coordinated accounts. Columns represent, from left to right: domain, the number of shares among coordinated accounts, and the factuality and political leaning scores based on MBFC.}
    \label{tab:telegram-domain-shares}
\end{table}

\smallskip

\textbf{$\mathbb{X}$/Twitter:} We identified a network of 19 coordinated accounts predominantly promoting content from right-leaning domains, as listed in Table~\ref{tab:twitter-domain-shares}.
A manual review of this $\mathbb{X}$ CoIA reveals that many of these coordinated accounts share similar profile descriptions, reflecting narratives tied to religious and conservative principles (see Table~\ref{tab:twitter-bios} in the Appendix). 
Additionally, an analysis of the messages (see Table~\ref{tab:5-messages-twitter} in the Appendix) shows a strong presence of politically partisan content, predominantly aligned with right-leaning ideologies.

\begin{table}[t]
    \footnotesize
    \centering
    \setlength{\tabcolsep}{8pt}
    \renewcommand{\arraystretch}{1.2}
    \begin{tabular}{c|c|c|c}
    \toprule
    \textbf{Domain} & \textbf{Shares} & \textbf{Factuality} & \textbf{Leaning} \\
    \midrule
    \href{www.foxnews.com}{foxnews.com} & 880 & Mixed & RIGHT \\
    \href{www.foxbusiness.com}{foxbusiness.com} & 13 & Mixed & RIGHT-CENTER \\
    \href{tmz.com}{tmz.com} & 5 & Mixed & RIGHT-CENTER \\
    \href{tmz.com}{conservativeinstitute.com} & 5 & Mostly & RIGHT \\
    \href{www.lifenews.com}{lifenews.com} & 2 & Low & FAR RIGHT \\
    \bottomrule
    \end{tabular}
    \caption{Top-5 domains shared in $\mathbb{X}$ by coordinated accounts. Columns represent, from left to right: domain, the number of shares among coordinated accounts, and the factuality and political leaning scores based on MBFC.}
    \label{tab:twitter-domain-shares}
\end{table}

\smallskip

\textbf{Facebook:} On Facebook, no well-known media outlets dominate the shared content. However, the most shared domain is a partisan news outlet, as suggested by the semantics of its title: \href{www.gorightnews.com}{gorightnews.com}. Additionally, there is a direct connection to a public activist known for political campaigns\footnote{\url{en.wikipedia.org/wiki/Peter_Boykin}} and their website: \href{www.peterboykin.com}{peterboykin.com}. These campaigns and related Facebook accounts primarily focus on the ``Gays for Trump'' movement, an American LGBTQ organization that advocates for U.S. President Donald Trump\footnote{\url{en.wikipedia.org/wiki/Gays_for_Trump}}. The top-engagement messages (see Table~\ref{tab:5-messages-facebook} in the Appendix) and bios (see Table~\ref{tab:facebook-bios} in the Appendix) indicate strong support for right-leaning ideologies.

\subsubsection{Cross-Platform Coordination.}

The analysis of the cross-platform network reveals distinct patterns.
The unified cross-platform network, displayed in Fig.~\ref{fig:cross-network}, highlights a giant component connecting coordinated accounts from both Telegram and $\mathbb{X}$, while the Facebook coordination network remains largely disconnected. A few noteworthy observations:

First, we observe that most cross-platform connections occur between $\mathbb{X}$ accounts and Telegram channels. Examining the biography descriptions of the top users by degree (see Table~\ref{tab:5-messages-cross} in the Appendix), we note a prominent presence of non-mainstream news outlets and conspiracy theories, such as the flat earth one.

Second, examining bridge nodes between the Telegram-coordinated cluster and the cross-platform giant component, we find highly influential channels with up to 24,000 subscribers. These channels often feature bios referencing free speech and religion, such as:
\begin{itemize}
    \item \textit{The FIGHT for FREEDOM - :Aron TRUTH Social}
    \item \textit{Q Reee-searchers Watchers}
    \item \textit{`Arise and shine, for your light has come, and the glory of the LORD rises upon you.' Isa. 60:1 WE RISE beyond the challenges of yesterday to become better stewards of tomorrow!}
\end{itemize}

Similarly, the bridge node between the $\mathbb{X}$-coordinated cluster and the cross-platform giant component is a highly influential account with 32,000 followers. The bio reads: \textit{``NO DM’S. Beautiful disaster. Self-proclaimed arbiter of great ideas. Here to annoy the dumb asses. NO LISTS!! \#Imvotingforafelon \#animallover''},
indicating a strong partisan flavor.

Finally, the largest component of the cross-platform coordination network is dominated by a mix of Telegram and $\mathbb{X}$ accounts promoting domains such as \texttt{magapac.com}, QAnon-related narratives (e.g., ``WWG1WGA''\footnote{\url{en.wiktionary.org/wiki/WWG1WGA}}), and accounts with partisan bios like:
\begin{itemize}
    \item \textit{We The Ultra Patriots is made up of patriotic Americans dedicated to exposing crimes against humanity, false flags, lies, and corruption. We The Ultra Patriots are...}
\end{itemize}

A closer look at the top shared domains within the coordinated network (see Table~\ref{tab:cross-domain-shares}) reveals that most are partisan, including \texttt{truthsocial.com}, an alternative non-mainstream social media platform\footnote{\url{en.wikipedia.org/wiki/Truth_Social}}. Additionally, six of the shared domains are Russian state-affiliated websites\footnote{\url{github.com/chan0park/VoynaSlov/tree/master}}, including \texttt{ruptly.tv} (2,117 shares), \texttt{rbc} (8 shares), \texttt{gazeta} (7 shares), \texttt{ruptly.video} (2 shares), \texttt{tv5} (1 share), and \texttt{mil} (1 share), further suggesting potential Russian information operations to interfere with the election discourse.
Notably, coordinated actors, just 1\% of observed Telegram accounts, shared over 2,000 messages linking to these media outlets—nearly 50\% of all links—demonstrating their outsized role in content amplification and none of these accounts is linked to any media organization.
Examining the posts and messages with the highest engagement (see Table~\ref{tab:5-messages-cross} in the Appendix), we observe the prevalence of ultra-MAGA narratives, conservative religious themes, and environmental news.

\begin{table}[t]
    \footnotesize
    \centering
    \setlength{\tabcolsep}{8pt}
    \renewcommand{\arraystretch}{1.2}
    \begin{tabular}{c|c|c|c}
    \toprule
    \textbf{Domain} & \textbf{Shares} & \textbf{Factuality} & \textbf{Leaning} \\
    \midrule
    \href{www.thegatewaypundit.com}{thegatewaypundit.com} & 23,513 & Very Low & EXTREME RIGHT \\
    \href{www.zerohedge.com}{zerohedge.com} & 8,331 & Low & RIGHT CONSPIRACY \\
    \href{www.truthsocial.com}{truthsocial.com} & 7,800 & NA & NA \\
    \href{www.nypost.com}{nypost.com} & 6,797 & Mixed & RIGHT-CENTER \\
    \href{www.theepochtimes.com}{theepochtimes.com} & 6,241 & Mixed & RIGHT \\
    \bottomrule
    \end{tabular}
    \caption{Top domains shared by cross-platform coordinated accounts. Columns represent, from left to right: domain, the number of shares among coordinated accounts, and the factuality and political leaning scores based on MBFC.}
    \label{tab:cross-domain-shares}
\end{table}

%%%%%%%%%%%%%%%%%%%%%%%%%%%%%%%%%%%%%%%%%%%%%%%%%%%%%%%%%
\begin{figure*}[h]
\centering
\begin{subfigure}{0.33\textwidth}
    \centering
    \includegraphics[width=\textwidth]{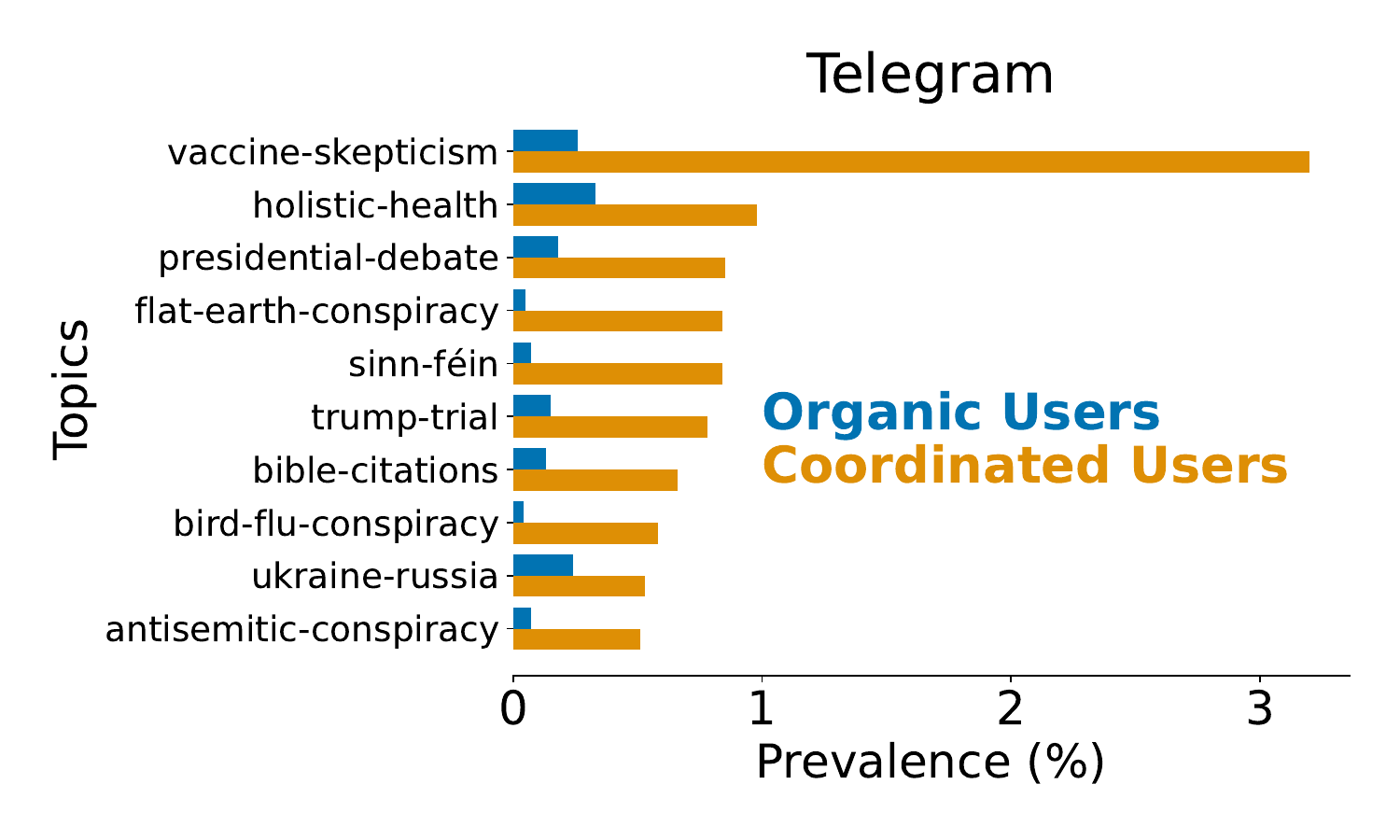}
    \caption{Telegram}
    \label{fig:telegram-topic-analysis}
\end{subfigure}%
\hfill
\begin{subfigure}{0.33\textwidth}
    \centering
    \includegraphics[width=\textwidth]{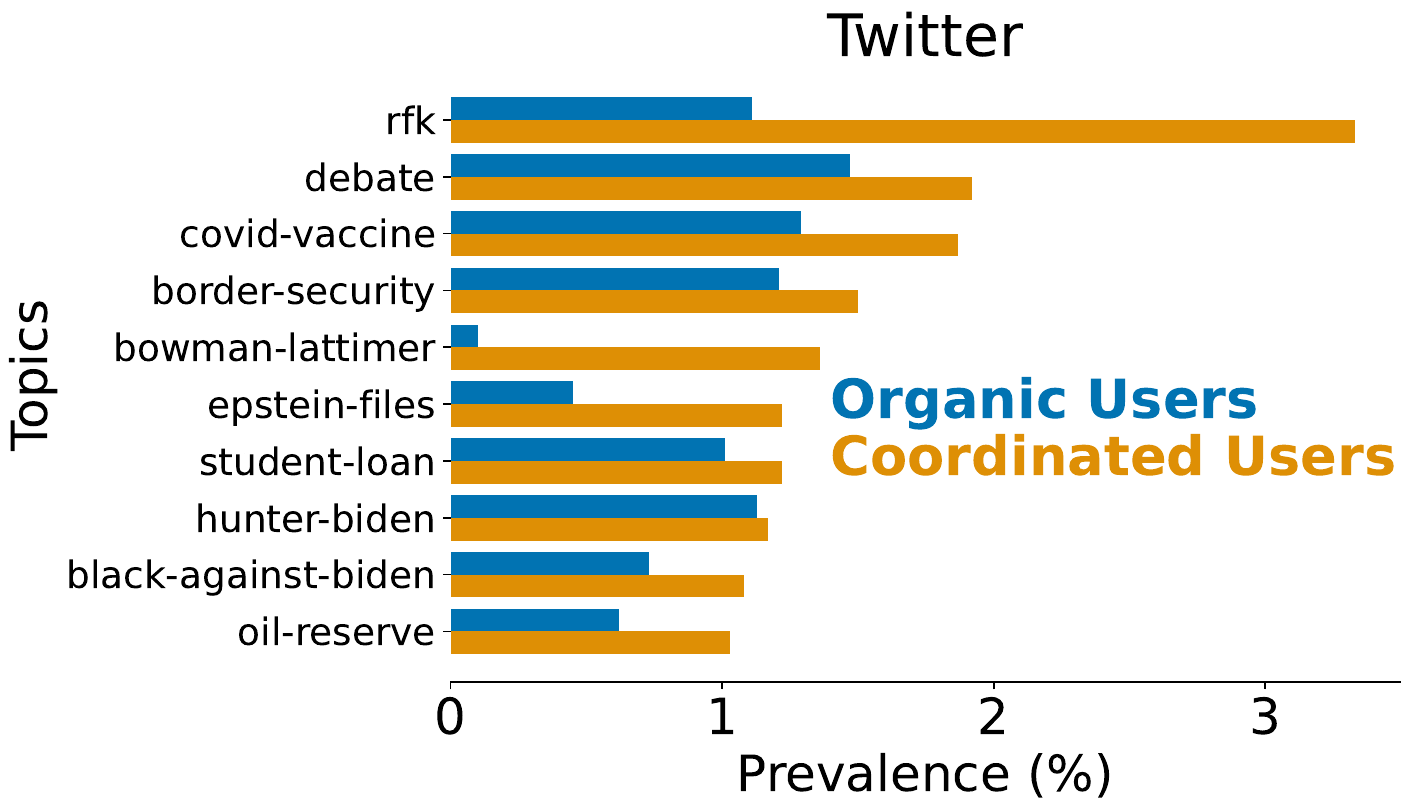}
    \caption{$\mathbb{X}$}
    \label{fig:twitter-topic-analysis}
\end{subfigure}%
\hfill
\begin{subfigure}{0.33\textwidth}
    \centering
    \includegraphics[width=\textwidth]{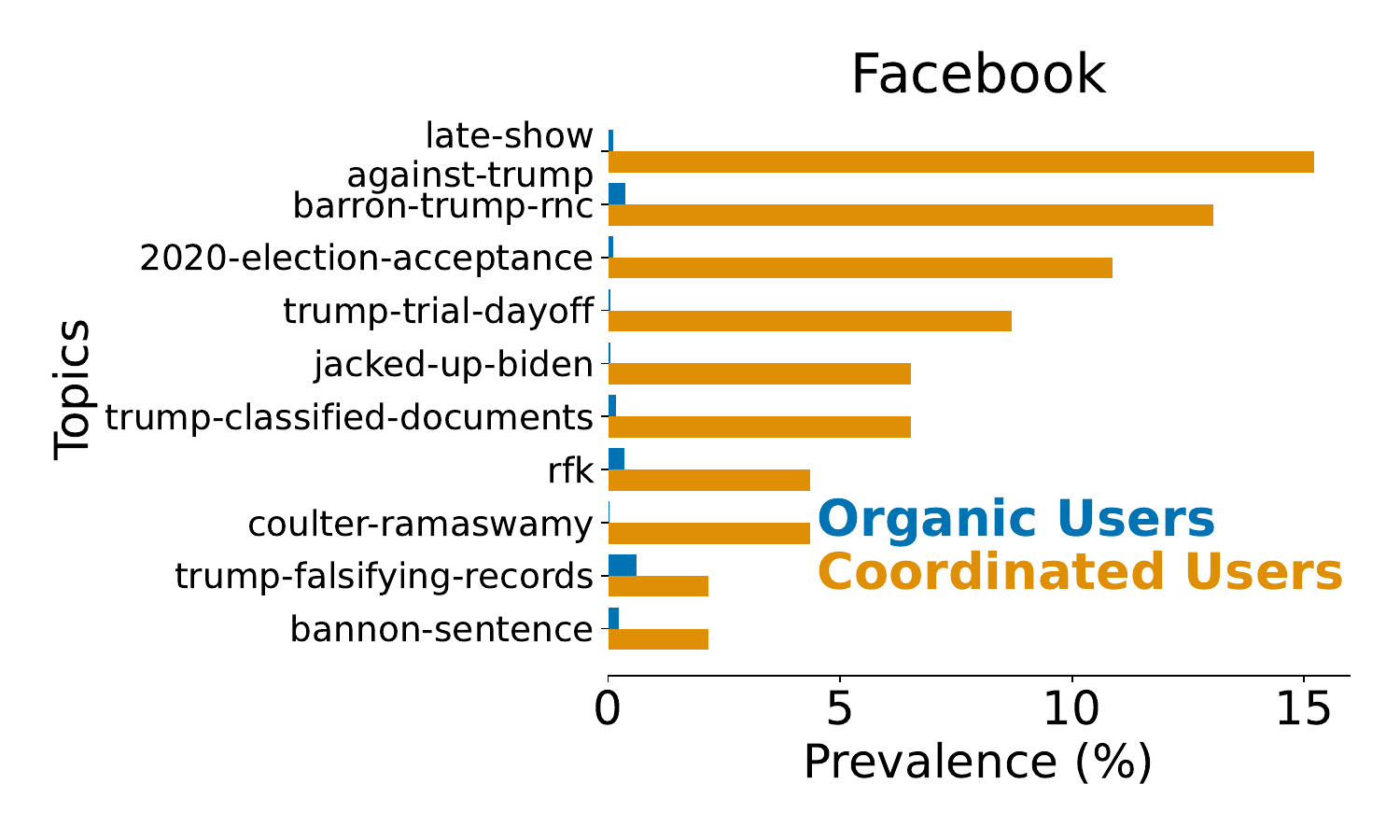}
    \caption{Facebook}
    \label{fig:facebook-topic-analysis}
\end{subfigure}
\caption{Content analysis of coordinated accounts on Telegram, $\mathbb{X}$, and Facebook. We report the top-10 most recurring topics and compare their prevalence against the organic discourse.}
\label{fig:topic-analysis}
\end{figure*}

\subsection{Detecting CoIA for Content Amplification (RQ2)}
In this section, we present our findings on the specific topics amplified by coordinated users within \textit{content amplification} campaigns. Following the method in \S\ref{sec:content-amplification}, we extract the Text Similarity Network and identify a seed set of coordinated users. We then apply BERTopic, as described in \S\ref{sec:topic-analysis}, to identify themes these users aim to promote and compare them to themes that emerge organically on each platform.
For each platform, we report the top-10 most prevalent topics within the coordinated cohort and compare their prevalence to that within the organic population. 
This approach highlights themes promoted by coordinated users, with large differences indicating topics characteristic of coordinated activity. In the remainder of this section, we detail our findings on coordination-driven content amplification across the three platforms.

\smallskip

\textbf{Telegram:}
Based on the significantly high similarity of their shared content, we identified 57 coordinated Telegram channels.
Among these channels, public health is a prominent topic of discussion, as can be seen in Figure~\ref{fig:telegram-topic-analysis}, with skepticism around the COVID-19 vaccine emerging as the most frequently discussed theme. 
Additionally, many messages promote alternative medicine, and there is notable discussion surrounding a conspiracy theory related to bird flu vaccination.
Telegram's coordinated channels also focus on strictly political discussions, particularly on the U.S. Presidential debate, the Russia-Ukraine conflict, and Donald Trump's legal issues. Interestingly, we observe discourse around immigration issues in Ireland, which appears to be echoed by accounts associated with the MAGA movement.
Notably, three out of the ten topics involve conspiracy theories related to flat-earth beliefs, bird flu, and antisemitism (see Table \ref{tab:topics_messages} in the Appendix).

We also assess the presence of QAnon-related keywords in content shared by both coordinated and organic channels. 
Approximately 1.66\% of content from coordinated users contains at least one QAnon-related keyword, compared to only 0.12\% of content from the organic population. 
This substantial difference is evident in Table~\ref{tab:keyword-prevalence}, where we list the top five keywords present in the coordinated group, also compared to keywords shared by organic channels.  Although the 57 coordinated channels make up only 0.36\% of the total number of Telegram channels in our dataset, they show a marked prominence, both in absolute and relative terms, in sharing QAnon-related keywords compared to organic channels. These findings indicate a propensity for Telegram CoIAs to disseminate conspiracies.
% theories---also linked to disinformation efforts during the 2020 U.S. election~\cite{ferrara2020characterizing}.

\begin{table}[t]
    \centering
    \begin{tabular}{lcc}
        \toprule
        \textbf{Keyword} & \multicolumn{2}{c}{\textbf{Count / Prevalence}} \\
        \cmidrule(lr){2-3}
         & \textbf{Coordinated} & \textbf{Organic} \\
        \midrule
        wwg1wga & 1260 (0.59\%) & 520 (0.05\%) \\
        plandemic & 619 (0.29\%) & 102 (0.01\%) \\
        adrenochrome & 448 (0.21\%) & 118 (0.01\%) \\
        qanon & 260 (0.12\%) & 107 (0.01\%) \\
        deepstate & 193 (0.09\%) & 87 (0.008\%) \\
        \bottomrule
    \end{tabular}
    \caption{QAnon-related keywords promoted by coordinated and organic channels on Telegram. Note that coordinated channels account for only 0.36\% of all Telegram channels. }
    \label{tab:keyword-prevalence}
\end{table}

\smallskip

\textbf{$\mathbb{X}$/Twitter:}
On $\mathbb{X}$/Twitter, we identified 221 coordinated users. 
Figure~\ref{fig:twitter-topic-analysis} displays the top 10 topics shared by these coordinated accounts, all of which pertain to social, economic, or political issues. 
The most prevalent topic---and one that differs significantly from those shared by the organic group---focuses on Independent candidate Robert F. Kennedy (RFK), with a subset of coordinated accounts actively promoting his candidacy in the U.S. Presidential Election.
Coordinated users also concentrate on the Democratic Primary race in Westchester County (labeled as ``bowman-lattimer'' in Figure~\ref{fig:twitter-topic-analysis}), where the most representative tweets endorse candidate Lattimer. This discussion appears to be framed around the anti-Israel stance of the incumbent, Rep. Jamaal Bowman.
The remaining topics revolve around polarizing issues such as COVID-19 vaccines, U.S. border security, and various scandals, including the Epstein files 
and Hunter Biden.
% and the second son of Joe Biden (Hunter Biden). 
Also, these users discuss perceived opposition within the Black community to President Biden, critiques of the student loan forgiveness policy, and concerns over the Biden administration's sale of oil reserves. 
It should be noted that the most representative tweets for each topic (as identified by BERTopic) were manually reviewed by three authors of this paper, who found no evidence of conspiratorial content (see Table \ref{tab:topics_tweets} in the Appendix). The most frequent QAnon-related keyword, “qanon,” appeared only five times.

\smallskip

\textbf{Facebook:}
We identify 16 coordinated users in the Facebook discourse. As shown in Figure~\ref{fig:facebook-topic-analysis}, most topics promoted by these users are left-leaning. A manual inspection reveals that all 16 accounts are associated with The Young Turks network\footnote{en.wikipedia.org/wiki/The\_Young\_Turks}, which MBFC classifies as left-leaning\footnote{mediabiasfactcheck.com/the-young-turks/}. Apart from a news item quoting a GOP representative accusing Biden of using ``supplements'' (i.e., ``jacked-up Biden''), we found no trace of misleading, conspiratorial, or inflammatory content.

%%%%%%%%%%%%%%%%%%%%%%%%%%%%%%%%%%%%%%%%%%%%%%%%%%%%%%%%%

\begin{figure*}[t]
    \centering
    \begin{tabular}{ccc}
        \includegraphics[width=0.31\textwidth]{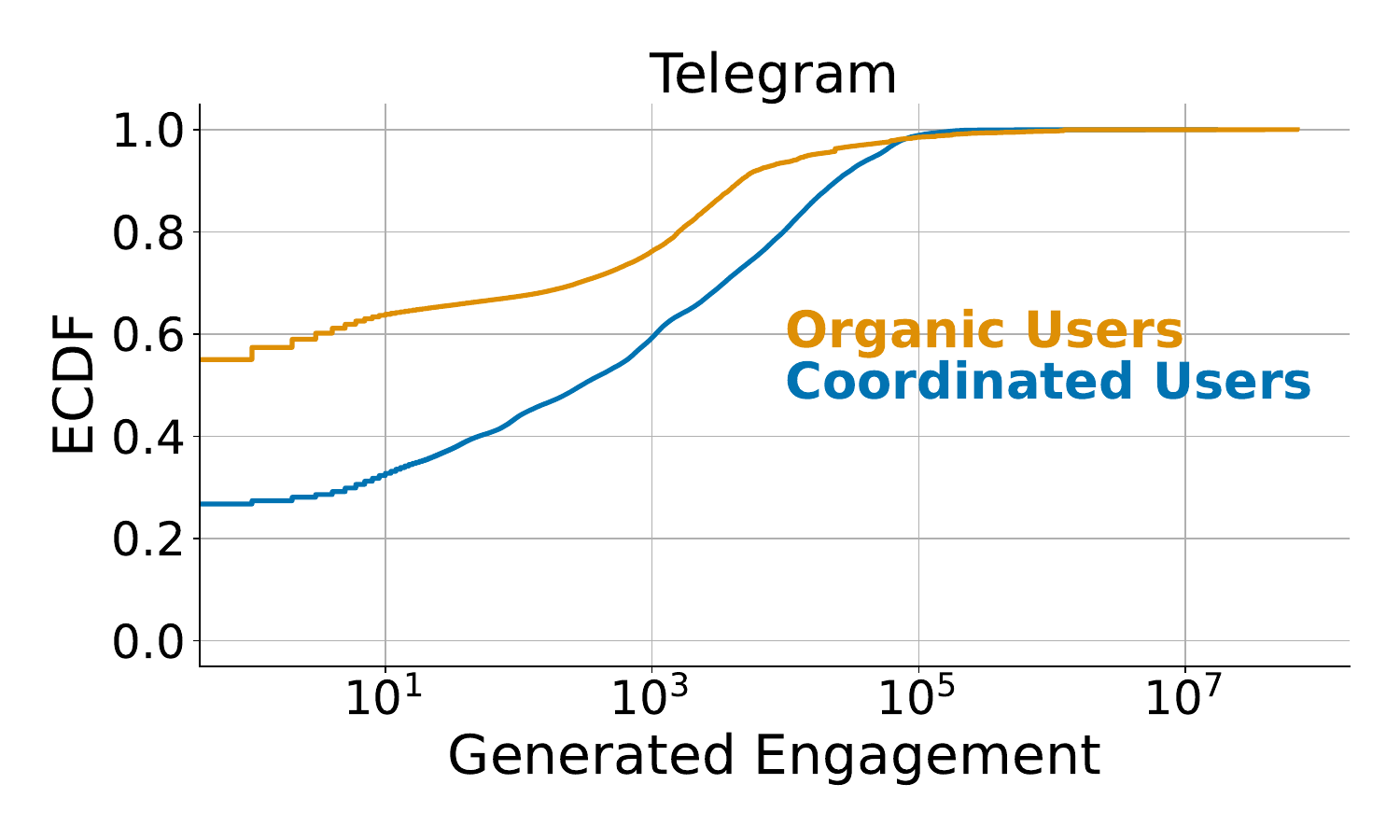}
         &
        \includegraphics[width=0.31\textwidth]{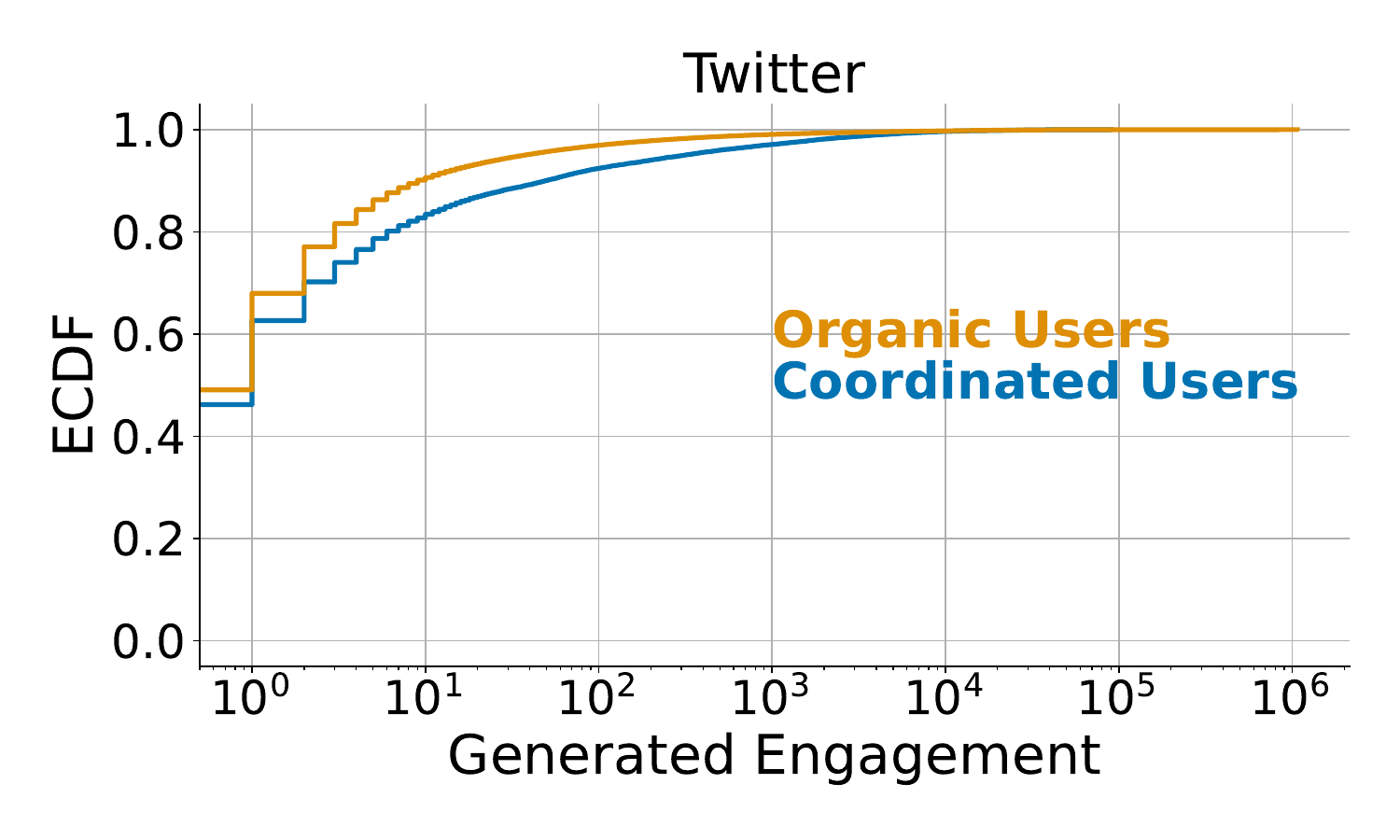} &
        \includegraphics[width=0.31\textwidth]{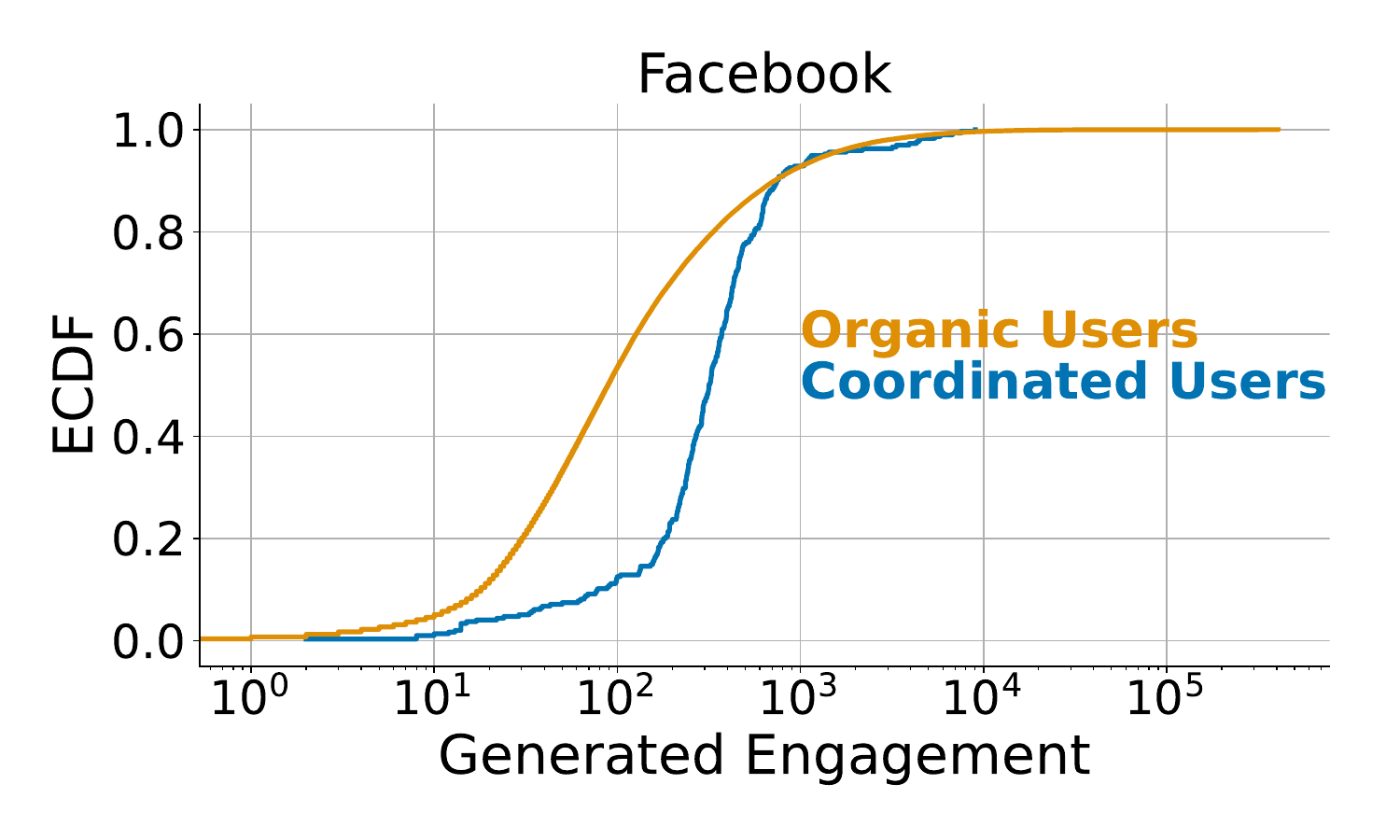} \\
    \end{tabular}
    \caption{Empirical Cumulative Distribution of total engagement, defined as the sum of user interactions across platforms, including reactions, comments, shares, replies, and views—generated by coordinated and organic users.}
    \label{fig:interactions}
\end{figure*}

\subsection{Generated Engagement and AIGC (RQ3)}
To comprehensively assess the impact of coordination strategies in the 2024 U.S. Election online debate, we analyze the level of engagement generated by coordinated actors and the extent of their reliance on AI-generated content (AIGC).
\begin{figure}[t]
    \centering    \includegraphics[width=.9\columnwidth]{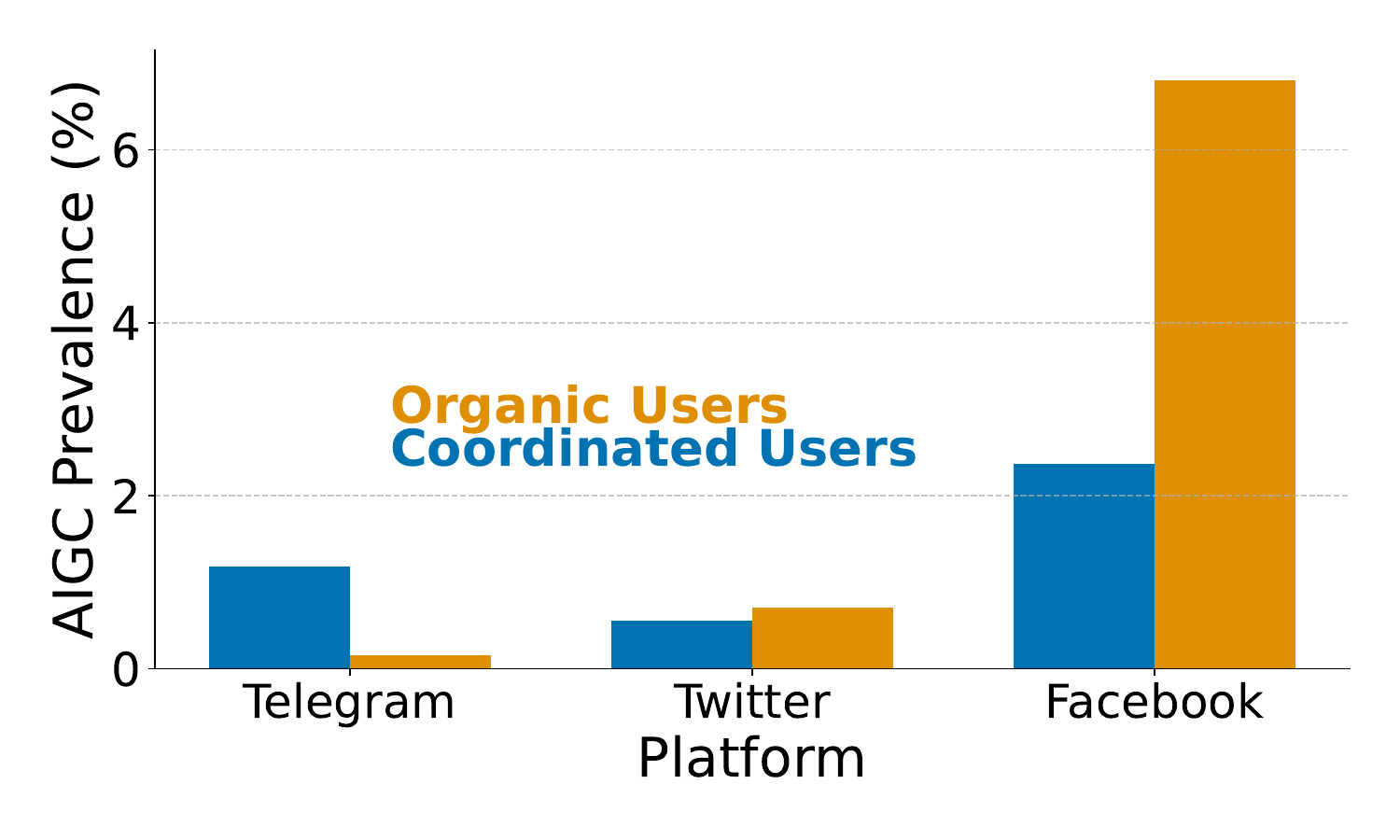}
    \caption{AIGC distribution across the three platforms, highlighting the prevalence of AI-generated posts among coordinated and organic users.}
    \label{fig:aigc-distribution}
\end{figure}
To maintain a consistent definition of engagement across platforms, we define it as the sum of likes, replies, quotes, and retweets for $\mathbb{X}$/Twitter, likes, comments, and shares for Facebook, and views, forwards, and replies for Telegram.

For each platform, we begin by merging the sets of coordinated actors identified through co-URL sharing (web domain promotion) and content similarity patterns (content amplification). We observe that the overlap between these two sets of coordinated users is modest, with overlap rates of 1.3\%, 0.5\%, and 0\% for Telegram, Twitter, and Facebook, respectively. 
This aligns with previous findings showing that different coordination strategies are employed to push diverse agendas through distinct coordinated networks~\cite{ng2022online, luceri2024unmasking}. This observation supports our decision to apply two complementary detection strategies. 
Aggregating coordinated actors from distinct CoIA networks, we identify 233 coordinated actors on Telegram, 764 on $\mathbb{X}$, and 25 on Facebook. The proportion of accounts engaged in CoIA is 1.49\% on Telegram, 0.43\% on $\mathbb{X}$, and 0.41\% on Facebook, consistent with prior research that employed conservative thresholds to detect coordinated activity~\cite{pacheco2021uncovering, luceri2024unmasking}.

% Table~\ref{fig:interactions} shows the distribution of engagement generated by coordinated and organic users across all platforms. In each case, coordinated users tend to generate less engagement than their organic counterparts. This is expected, as the organic group includes major media outlets and influencers that significantly boost overall engagement. However, it is noteworthy that the engagement from coordinated actors is substantial, and particularly on $\mathbb{X}$ and Facebook, the engagement distributions of coordinated and organic users are quite similar.
Figure~\ref{fig:interactions} shows the distribution of engagement generated by coordinated and organic users across all platforms. 
On Telegram, more than half of the content generated by organic accounts receives no engagement, while the median engagement for coordinated accounts is 297—highlighting the ability of coordinated users to consistently amplify their content.
On Twitter/$\mathbb{X}$, the engagement distribution is more similar across the two groups, with coordinated users generating slightly more engagement, as suggested by the organic curve consistently lying above the coordinated one. 
On Facebook, the median engagement of coordinated users (317) substantially exceeds that of organic users (89). This may result from reciprocal interactions within the coordinated cohort, a strategic effort by coordinated users to stimulate organic engagement, or a combination of both. In contrast, many organic users may have little or no interest in maximizing engagement.
% In each case, coordinated users tend to generate less engagement than their organic counterparts. This is expected, as the organic group includes major media outlets and influencers that significantly boost overall engagement. However, it is noteworthy that the engagement from coordinated actors is substantial, and particularly on $\mathbb{X}$ and Facebook, the engagement distributions of coordinated and organic users are quite similar.

We also assess the prevalence of AIGC that was shared across the three platforms by both coordinated and organic users. These results are shown in Figure~\ref{fig:aigc-distribution}.
Facebook is the social media platform where most AIGC is produced by both coordinated and organic accounts. 
While the AIGC prevalence is limited, we observe a striking difference in the activity between coordinated and organic users on Telegram and Facebook, whereas it is quite balanced on $\mathbb{X}$. Interestingly, AI-generated content is predominantly diffused by coordinated actors on Telegram, while an opposite trend is observed on Facebook.
% \clearpage

\section{Conclusions}
% !TEX root =  ./main.tex
\label{sec:conclusions}
This paper introduces a novel, network-based framework for detecting coordinated inauthentic activity (CoIA) across multiple social media platforms. Unlike most traditional methods that focus on isolated platform-specific analyses, our approach emphasizes cross-platform behaviors by constructing similarity networks that capture community-wide patterns of content sharing. By prioritizing both intra- and cross-platform connections, we are able to detect subtle yet significant coordination signals that remain hidden when platforms are analyzed in isolation.

Our unsupervised, network-based methodology allows us to identify a range of influence campaigns aimed at directing traffic toward specific narratives and domains. These campaigns frequently promote content that is partisan, low-credibility, and conspiratorial in nature. The model not only detects domestic actors but also reveals the systematic promotion of foreign-affiliated media, particularly Russian state-sponsored outlets, across platforms like Telegram and $\mathbb{X}$. This evidence points to a coordinated effort to amplify certain messages and domains across distinct platforms
% , particularly in the context of the 2024 U.S. Election, 
underscoring the necessity of a unified, cross-platform approach to tackle such influence operations \cite{gatta2023interconnected}.

\spara{Limitations.}
While our study demonstrates substantial strengths, a few considerations remain. 
First, although the model leverages data from Telegram, Facebook, and $\mathbb{X}$, the specific timeframe under observation and platform-specific mechanisms may influence the broader applicability of our model and the generalizability of our findings to other scenarios and social media networks. 
Second, combining similarity networks with differing similarity distributions across platforms may introduce biases in the overall detection of coordination, potentially favoring certain platforms. 
% This approach emphasizes adaptability to platform-specific dynamics, though it may pose challenges for direct cross-platform comparisons of coordination strength.
Finally, this work does not examine multimedia content, thereby overlooking coordination across different data modalities and the potential use of generative AI techniques for producing synthetic media.

Future work should address these limitations by incorporating sophisticated statistical models to assess the significance of coordination patterns, thereby providing a more robust and statistically sound methodology to ensure that identified coordination signals are not merely artifacts of underlying data distributions. Expanding the framework to encompass additional platforms, data modalities, and a broader range of behavioral signals would further strengthen its capability to detect diverse forms of coordination.

\newpage

\bibliographystyle{ACM-Reference-Format}
\balance
\bibliography{references}

\appendix
\renewcommand\thefigure{\thesection.\arabic{figure}}
\renewcommand\thetable{\thesection.\arabic{table}}
\setcounter{figure}{0}
\setcounter{table}{0}
% !TEX root =  ../main.tex
\section*{About the Team}
The 2024 Election Integrity Initiative is led by Emilio Ferrara and Luca Luceri and carried out by a collective of USC students and volunteers whose contributions are instrumental to enable these studies. The authors are indebted to Srilatha Dama and Zhengan Pao for their help in bootstrapping this data collection. The authors are also grateful to the following HUMANS Lab's members for their tireless efforts on this project: Ashwin Balasubramanian, Leonardo Blas, Charles 'Duke' Bickham, Keith Burghardt, Sneha Chawan, Vishal Reddy Chintham, Eun Cheol Choi, Priyanka Dey, Isabel Epistelomogi, Saborni Kundu, Grace Li, Richard Peng, Gabriela Pinto, Jinhu Qi, Ameen Qureshi, Tanishq Salkar, Kashish Atit Shah, Reuben Varghese, Siyi Zhou.
Previous memos:
\cite{ferrara2024charting,ferrara2024risks,pinto2024tracking,minici2024uncovering,blas2024unearthing,balasubramanian2024public}.

\section{Dataset Collection}

\noindent{\textbf{Keywords.}}
Joe Biden, Donald Trump, 2024 US Elections, US Elections, 2024 Elections, 2024 Presidential Elections, Biden, Joe Biden, Joseph Biden, Biden2024, Donald Trump, Trump2024, trumpsupporters, trumptrain, republicansoftiktok, conservative, MAGA, KAG, GOP, CPAC, Nikki Haley, Ron DeSantis , RNC, democratsoftiktok, thedemocrats, DNC, Kamala Harris, Marianne Williamson, Dean Phillips, williamson2024, phillips2024, Democratic party, Republican party, Third Party, Green Party, Independent Party, No Labels, RFK Jr, Roberty F. Kennedy Jr. , Jill Stein, Cornel West, ultramaga, voteblue2024, letsgobrandon, bidenharris2024, makeamericagreatagain, Vivek Ramaswamy, JD Vance, Assassination, Tim Walz, WWG1WGA.

\noindent{\textbf{Additional dataset statistics.}} We present the distribution of the number of posts per user and post's length for each platform.

\label{app:dataset}
\begin{table*}
    \centering
    \begin{tabular}{cc}
        % First row
        \includegraphics[width=1.\columnwidth]{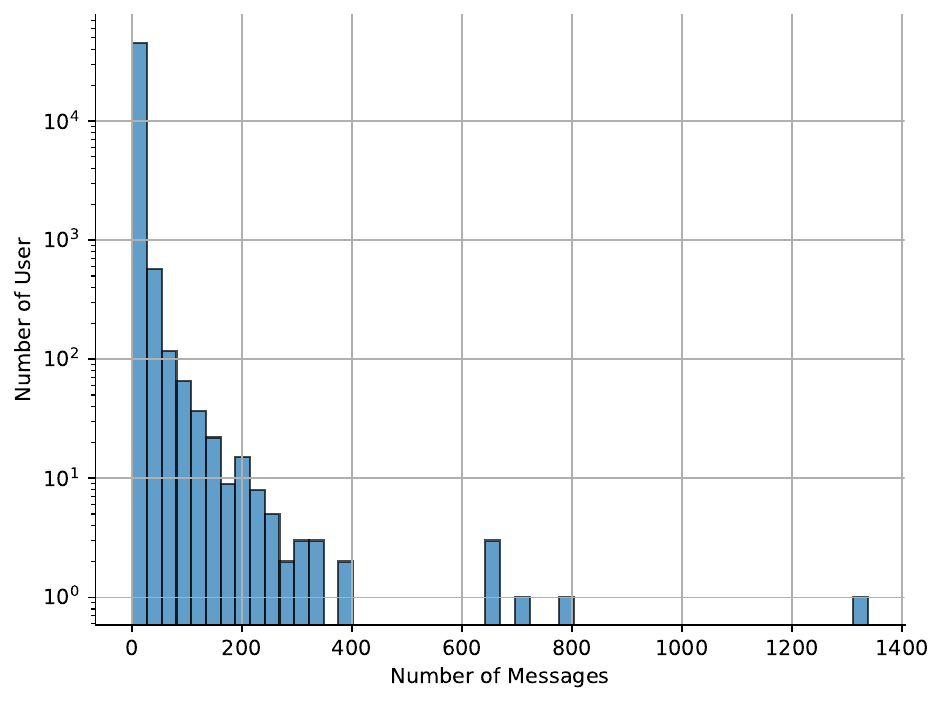} &
        \includegraphics[width=1.\columnwidth]{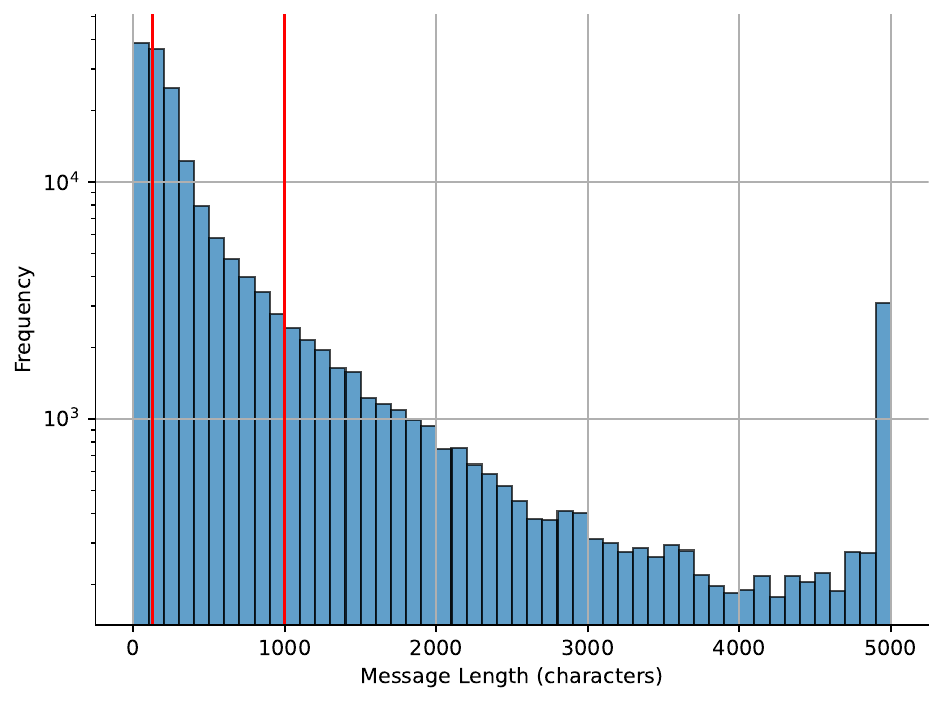} \\
        \textbf{(a) Facebook posts} & \textbf{(b) Facebook characters} \\
        
        % Second row
        \includegraphics[width=1.\columnwidth]{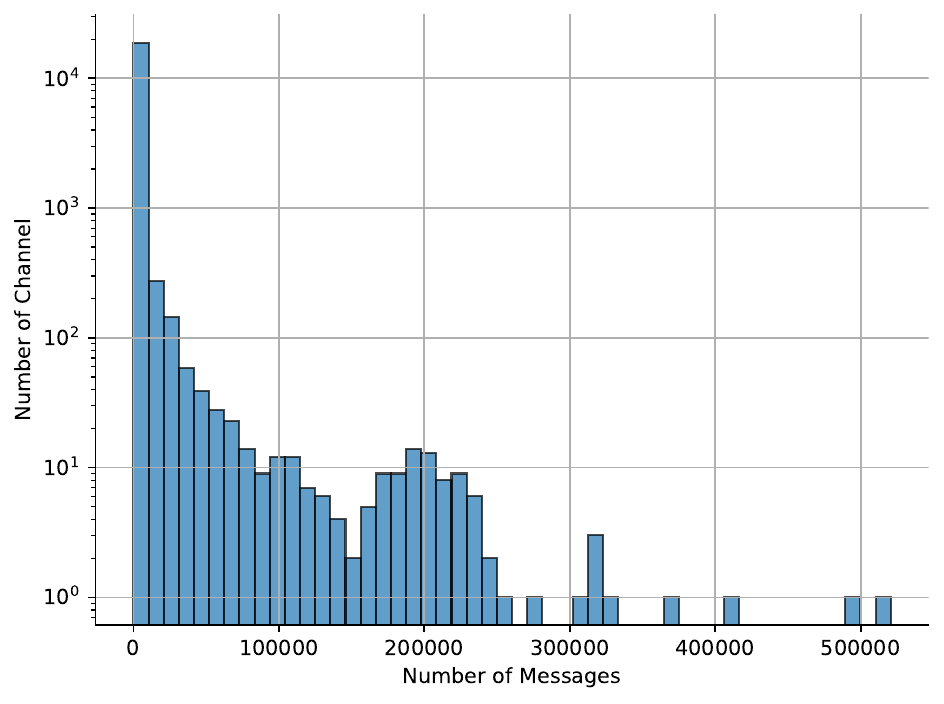} &
        \includegraphics[width=1.\columnwidth]{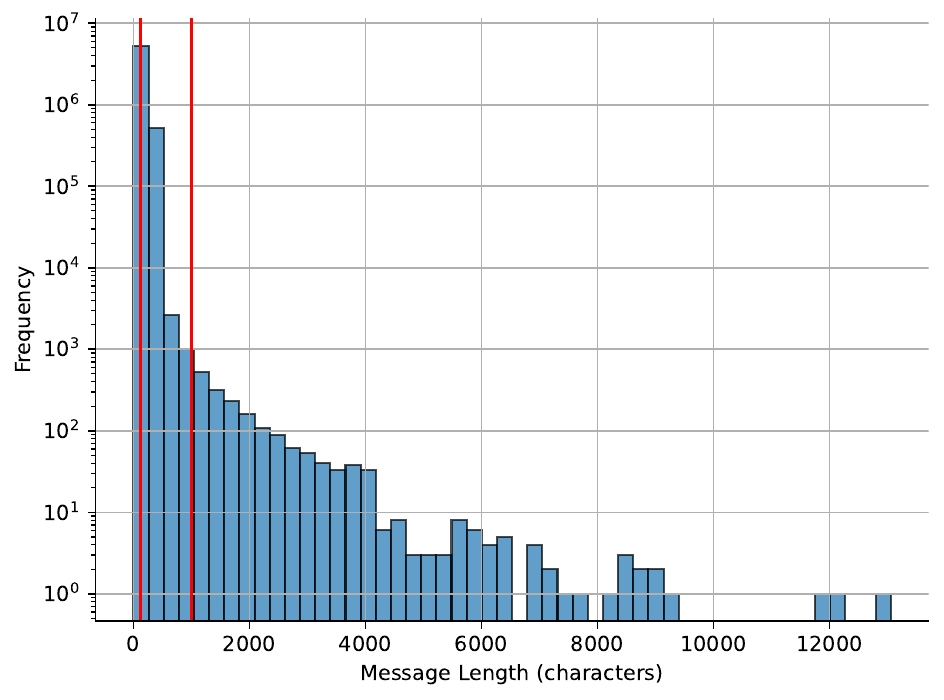} \\
        \textbf{(c) Telegram posts} & \textbf{(d) $\mathbb{X}$ characters} \\
        
        % Third row
        \includegraphics[width=1.\columnwidth]{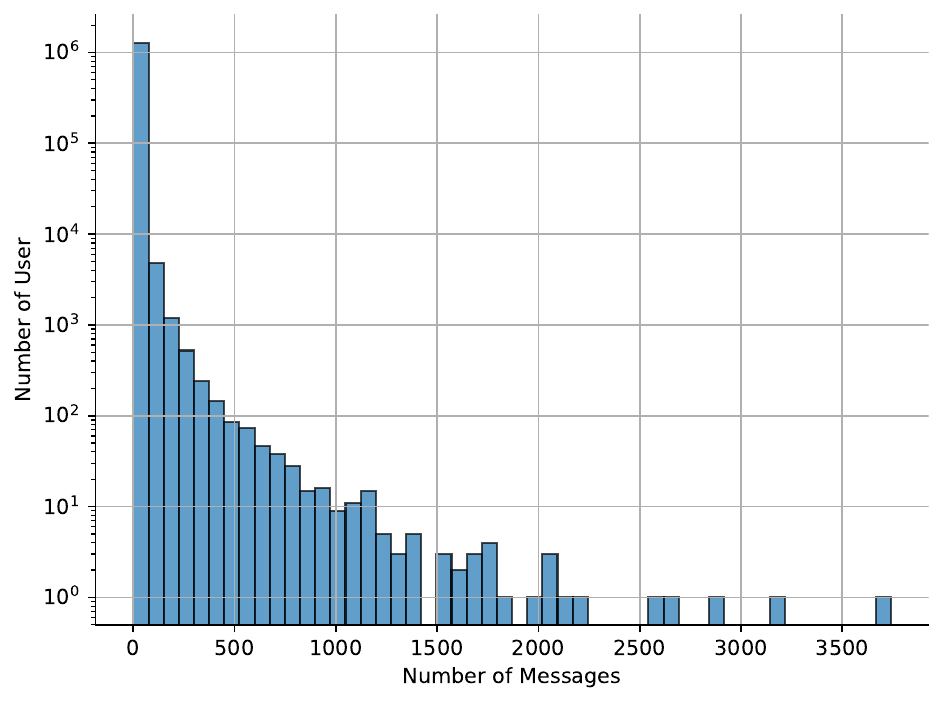} &
        \includegraphics[width=1.\columnwidth]{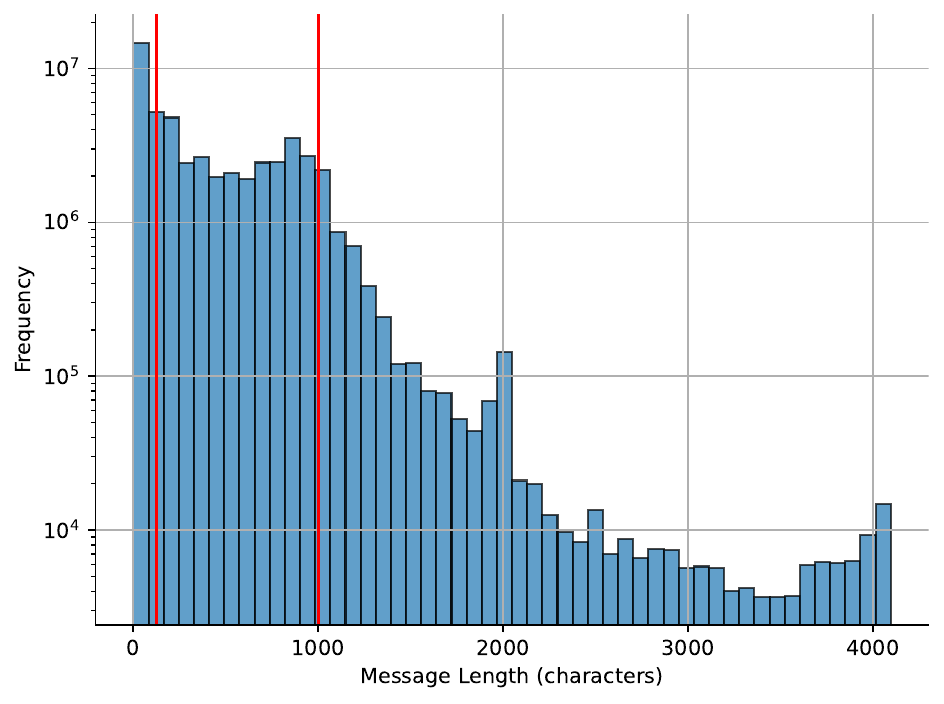} \\
        \textbf{(e) $\mathbb{X}$ posts} & \textbf{(f) Telegram characters} \\
    \end{tabular}
    \caption{Distributions of posts and characters across platforms.}
    \label{fig:posts-characters-distrib}
\end{table*}

\section{AI Detection Classifier Validation}
We utilized the AI-generated text detection classifier from~\cite{dmonte2024classifying}, originally trained to identify AI-generated content in tweets. To extend the applicability of this classifier to other platforms, we constructed an external validation set using diverse, older datasets. Specifically, we built upon four datasets: \texttt{Twitter-2010}\footnote{\url{https://archive.org/details/twitter_cikm_2010}}, \texttt{Facebook}\footnote{\url{https://www.kaggle.com/datasets/sheenabatra/facebook-data}}, and \texttt{Telegram}.

From each of these datasets, we sampled 500 texts ranging in length from 125 to 1000 characters. Given the release years of these datasets, we assume that all texts are non-AI generated. To create the AI-generated class, we used GPT-4o\footnote{\url{https://openai.com/index/hello-gpt-4o/}} and Llama 3.1 3B Instruct\footnote{\url{https://huggingface.co/meta-llama/Meta-Llama-3-8B-Instruct}}, prompting them to generate new texts with similar topics and lengths as the original non-AI texts. We performed this generation for each non-AI text and sampled 250 AI-generated texts from GPT and 250 from Llama.

Thus, for each platform, the validation dataset consists of 500 non-AI generated texts paired with 500 AI-generated texts: 250 from GPT and 250 from Llama, with both AI-generated sets matching the original texts in topic and length.

To further enhance the generalizability of the validation dataset, we incorporated non-AI generated content from additional sources:  Reddit comments\footnote{\url{https://zissou.infosci.cornell.edu/convokit/datasets/reddit-coarse-discourse-corpus/}}, IMDB reviews\footnote{\url{https://ai.stanford.edu/~amaas/data/sentiment/}}, movies corpus\footnote{\url{https://zissou.infosci.cornell.edu/convokit/datasets/movie-corpus/}}, and Yelp 2013\footnote{\url{https://www.yelp.com/dataset/download}}. AI-generated content for these datasets was sourced from tweetHunter\footnote{\url{https://tweethunter.io/}} and GPT, which was prompted to generate texts on specific topics such as American football, climate change, the Soccer World Cup, the Gaza conflict, U.S. politics, and vaccines.

The validation results are displayed in Table~\ref{table:precision_recall}.

\begin{table}[ht]
\centering
\begin{tabular}{c|c|c}
\hline
\textbf{Platform} & \textbf{Precision} & \textbf{Recall} \\
\hline
General & 0.9064 & 0.5927 \\
\midrule
\midrule
Facebook & 0.8723 & 0.6686 \\
$\mathbb{X}$/Twitter & 0.8935 & 0.5613 \\
Telegram & 0.9598 & 0.5956 \\
\midrule
Reddit & 0.895 & 0.5368 \\
Additional set & 0.9697 & 0.5 \\
\midrule
\end{tabular}
\caption{Precision and Recall for each platform}
\label{table:precision_recall}
\end{table}
\label{app:ai-detection}

\section{Grid Search for Similarity and Centrality Thresholds}
\begin{figure*}[hb!]
    \centering
    \begin{tabular}{cccc}
        % First row
        \begin{subfigure}{0.33\textwidth}
            \centering
            \includegraphics[width=\textwidth]{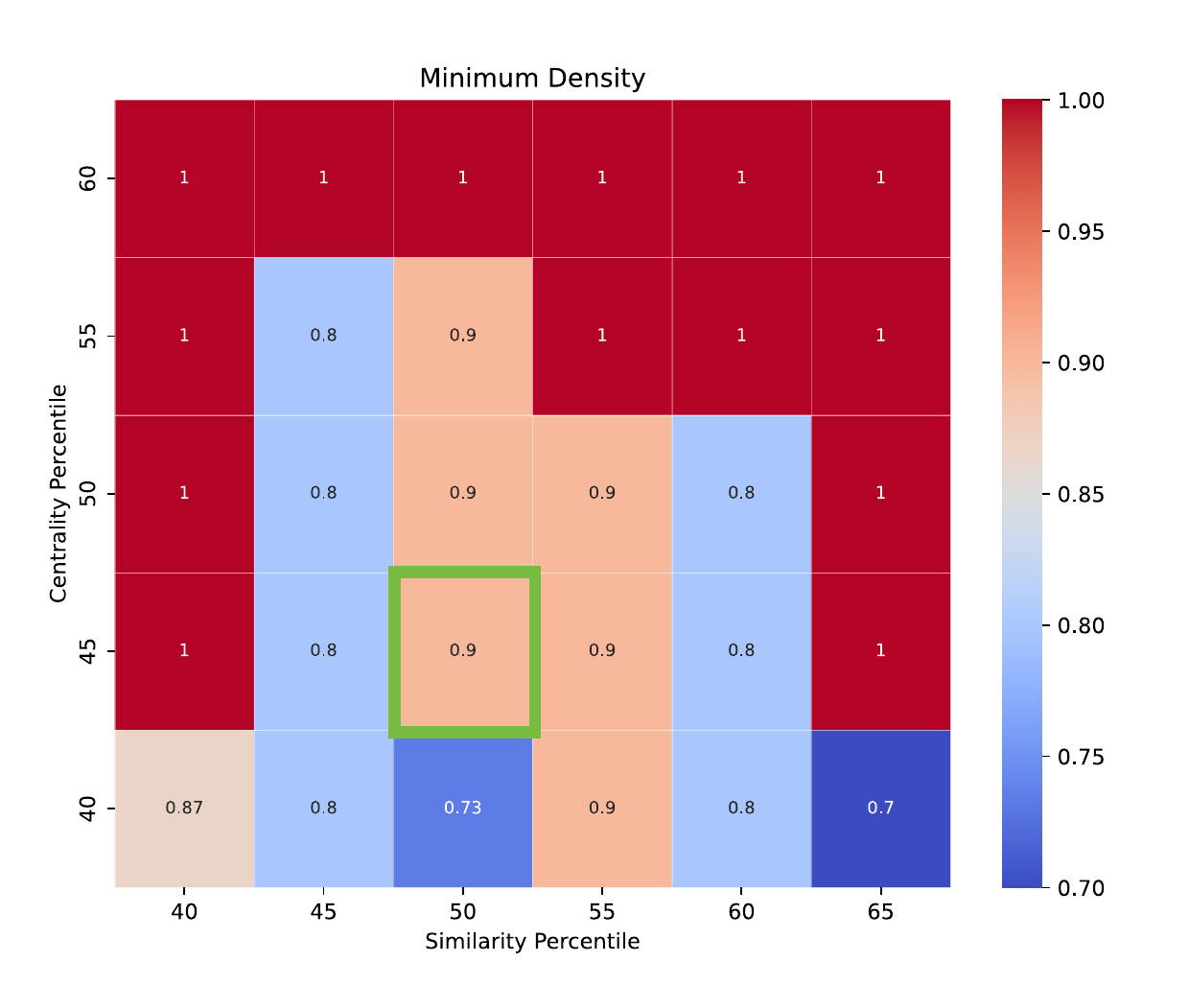}
            \caption{Facebook.}
            \label{fig:facebook-posts-distrib}
        \end{subfigure} &
        \begin{subfigure}{0.33\textwidth}
            \centering
            \includegraphics[width=\textwidth]{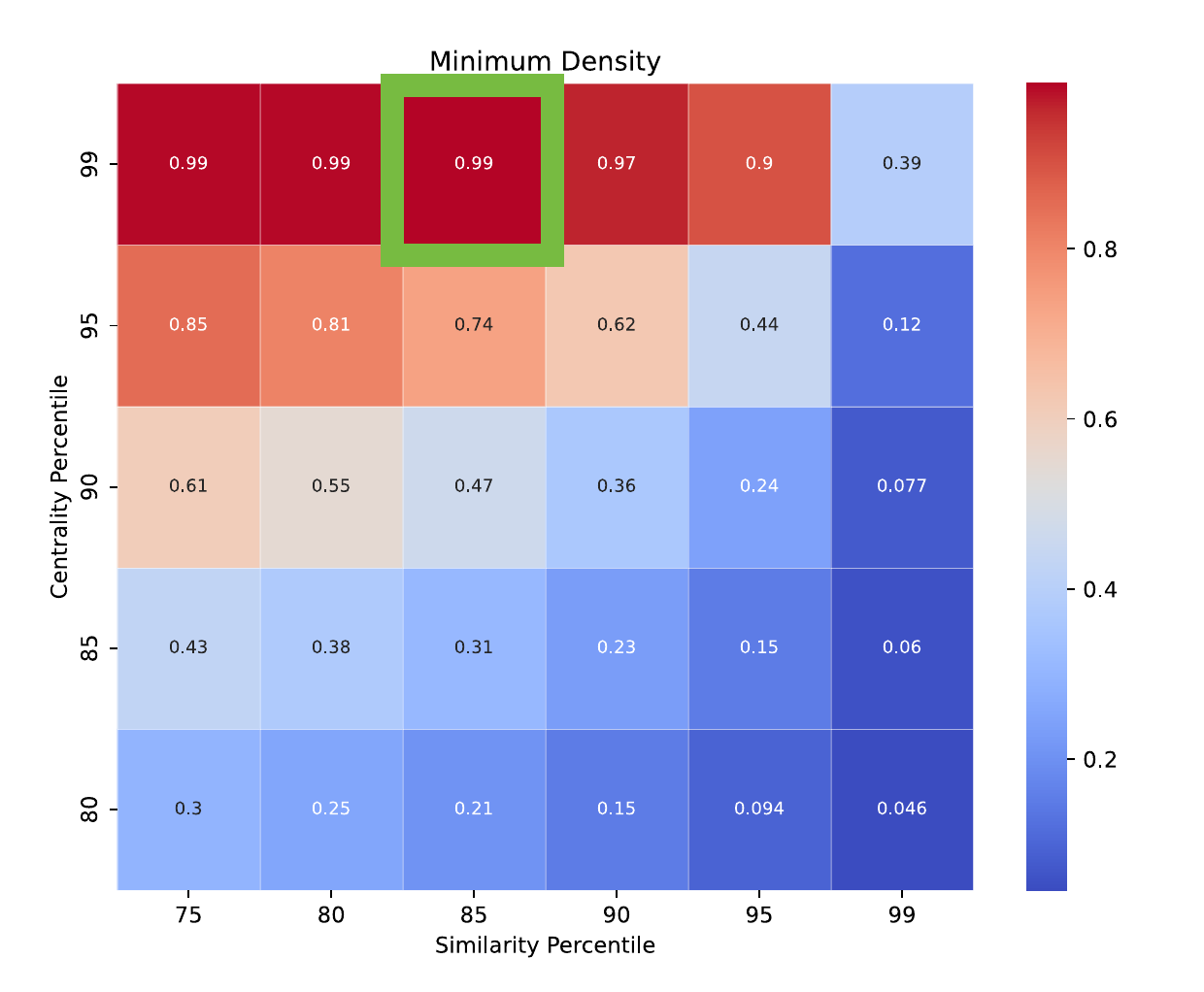}
            \caption{$\mathbb{X}$.}
            \label{fig:twitter-posts-distrib}
        \end{subfigure} &
        \begin{subfigure}{0.33\textwidth}
            \centering
            \includegraphics[width=\textwidth]{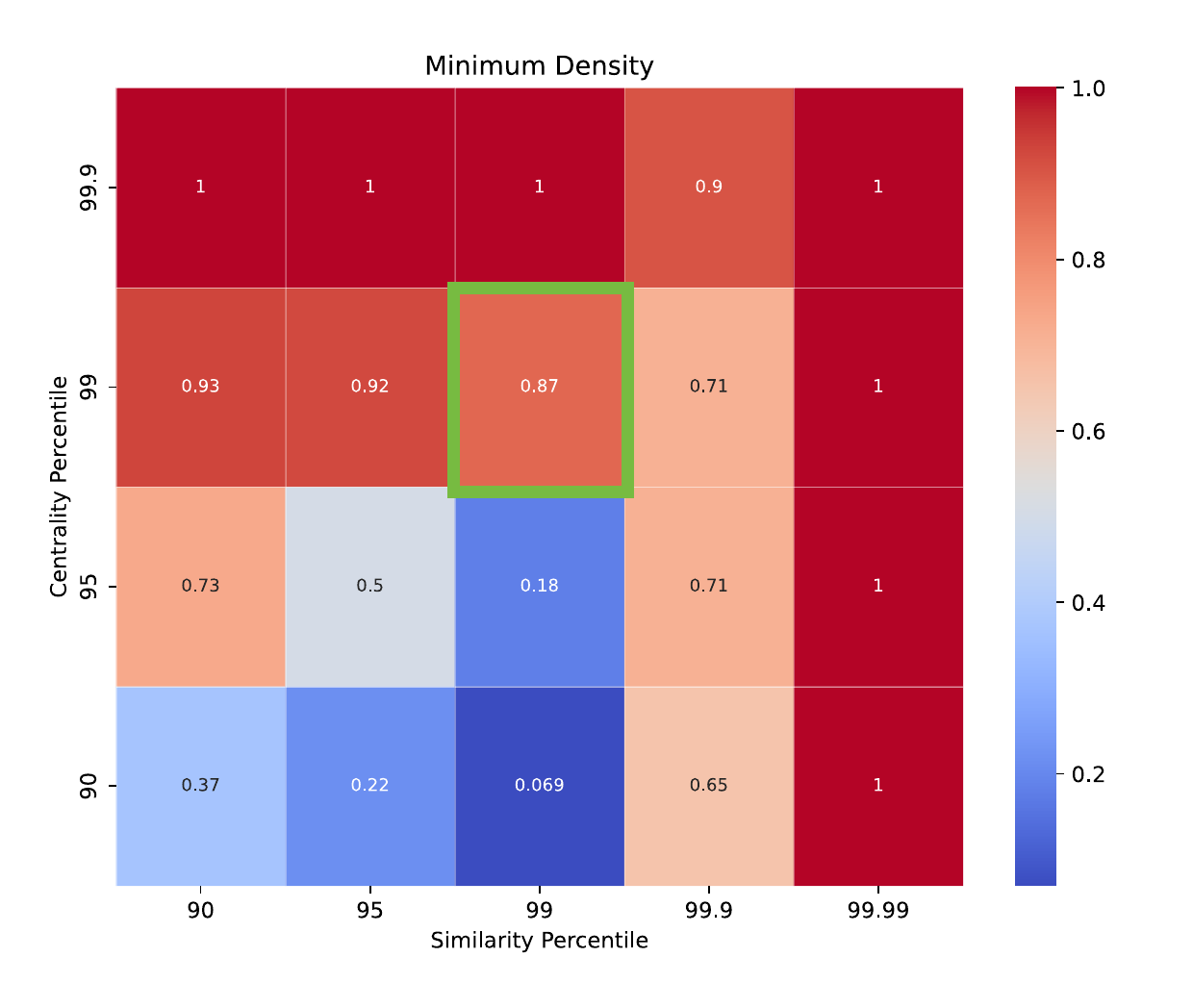}
            \caption{Telegram.}
            \label{fig:telegram-posts-distrib}
        \end{subfigure} \\
    \end{tabular}
    \caption{Thresholds grid search for filtering similarity graph: x-axis edge similarity quantile, y-axis node centrality quantile. z-axis is the minimum graph density across all connected components of the filtered co-url similarity graph. The green square corresponds to the selected thresholds.}
    \label{fig:heatmaps-grid}
\end{figure*}

To identify coordinated accounts in the co-URL similarity network, we filter edges and nodes based on cosine similarity and eigenvector centrality. Two nodes are connected if they co-share URLs, and the strength of these connections is represented by the cosine similarity between their TF-IDF vectors across the space of all unique URLs.
We employ two filtering techniques commonly found in the literature: edge filtering, based on cosine similarity, and node filtering, based on eigenvector centrality. The thresholds for these filters are determined by the percentiles of the distributions of edge similarity and node centrality.
Assuming that coordinated accounts manifest as dense components in the similarity graph, we use the density of each connected component as a key indicator of coordination. To ensure robustness, we adopt a conservative approach, using the minimum density across all connected components after filtering as a comprehensive quality measure of the graph.
Figure~\ref{fig:heatmaps-grid} presents the results of our grid search across these parameters. The x-axis represents the percentile of edge similarity, while the y-axis corresponds to the percentile of eigenvector centrality. The z-axis indicates the minimum density of the similarity graph after filtering based on the selected x-y percentiles.
For each platform, we identified the optimal thresholds by observing sharp changes in minimum density and high overall density values. These threshold combinations define the coordinated accounts on each platform. Specifically:
\begin{itemize}
    \item For Facebook, we selected (50\%, 45\%) as the minimum density increased from 0.73 and 0.80 to 0.90.
    \item For $\mathbb{X}$, we chose (85\%, 99\%), consistent with previous studies, leading to an increase from 0.74 to 0.99.
    \item For Telegram, we used (99\%, 99\%), focusing on a significant jump from 0.18 to 0.87.
\end{itemize}

Other threshold combinations were evaluated qualitatively, yielding similar results. More quantitative methods for threshold selection are left for future work.

\smallskip

\noindent{\textbf{Performance Evaluation.}}
To validate our methodology, we used the supervised Information Operations (IO) dataset from~\cite{seckin2024labeled}, which includes labeled users identified as coordinated inauthentic actors across 26 IO campaigns, along with a control group of organic users.
Our method achieves high precision (above 0.9) but low recall (around 0.1), aligning with the framework’s design and the objectives of our analysis.

\clearpage

\begin{figure*}
    \centering
    \includegraphics[width=0.7\textwidth]{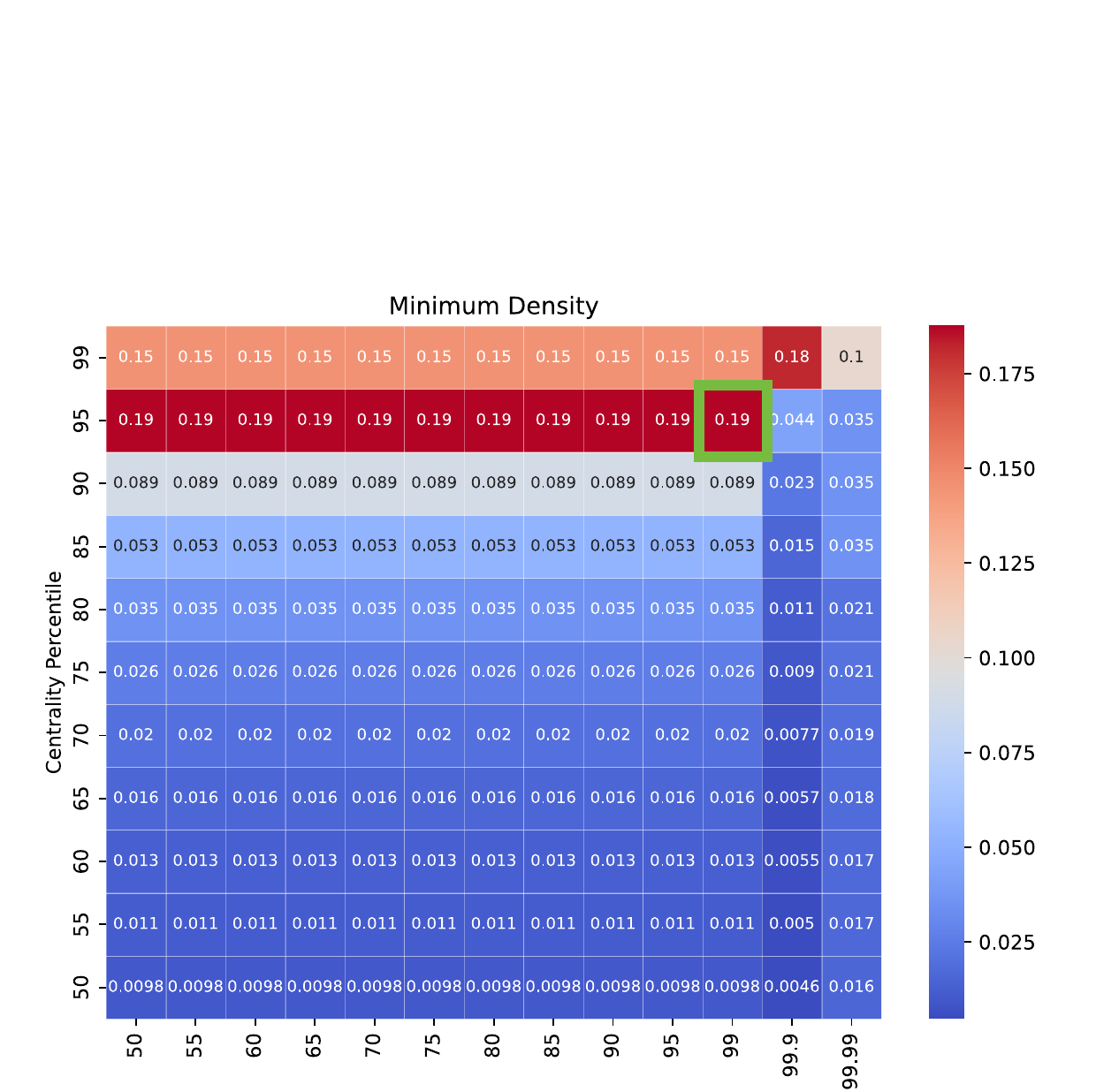}
    \caption{Thresholds grid search for filtering the Cross-platform similarity graph: x-axis edge similarity quantile, y-axis node centrality quantile. z-axis is the minimum graph density across all connected components of the filtered co-url similarity graph. The green square corresponds to the selected thresholds.}
    \label{fig:heatmap-cross}
\end{figure*}

\section{Characterization of coordinated users}
We report the bios of the top 5 coordinated accounts by degree, at the time of writing, and the top 5 messages by total engagement written by coordinated accounts in each platform.

\begin{table}[t]
    \scriptsize
    \centering
    \setlength{\tabcolsep}{8pt}
    \renewcommand{\arraystretch}{1.2}
    \begin{tabular}{c}
    \toprule
    \textbf{Bio}  \\
    \midrule
    CV Lies \& 5G Dangers - Discussions \\
    \midrule
    Scammers may try to impersonate Real World News Channel.\\
    Don’t trust any messages you get from them.\\
    They are not us. Block and report them immediately. \\
    \midrule
    Covering News, Military information, across Wiltshire \& the \\
    Southwest areas and for connecting people together uk\_flag \\
    \midrule
    Credence - Breaking News \\
    \midrule
    Truth. Faith. Freedom.\\
    \bottomrule
    \end{tabular}
    \caption{Current bios of top-5 coordinated channels in Telegram sorted by degree.}
    \label{tab:telegram-bios}
\end{table}

\begin{table}[t]
    \scriptsize
    \centering
    \setlength{\tabcolsep}{8pt}
    \renewcommand{\arraystretch}{1.2}
    \begin{tabular}{c}
    \toprule
    \textbf{Bio}  \\
    \midrule
    Have integrity. Be Just. Precinct Chairman. \\Christian. Pro-Life. Conservative Political Activist. \#CruzCrew\\
    \midrule
    NO DM’S. Beautiful disaster. \\Self proclaimed arbiter of great ideas. \\Here to annoy the dumb asses. \\NO LIST!! \#Imvotingforafelon \#animallover\\
    \midrule
CHRISTIAN, ULTRA MAGA, PATRIOT, CONSERVATIVE, \\2A, FJB, TRUMP WON, TRUMP2016, TRUMP2020, TRUMP2024,\\ ELON MUSK, YELLOWSTONE, DALLAS COWBOYS \\
    \midrule
    just a opinionated old cowgirl from WY  \\Traditional work ethic, traditional values \\
    \midrule
    studied spirituality under a minister study 15 years worked 30+years.\\ I write as I see them, past present and maybe future. \\don’t expect you to agree but think\\
    \bottomrule
    \end{tabular}
    \caption{Current bios of top-5 coordinated accounts in $\mathbb{X}$ sorted by degree.}
    \label{tab:twitter-bios}
\end{table}

\begin{table}[t]
    \scriptsize
    \centering
    \setlength{\tabcolsep}{8pt}
    \renewcommand{\arraystretch}{1.2}
    \begin{tabular}{c}
    \toprule
    \textbf{Bio}  \\
    \midrule
    \#GayRightNews
Bringing You Gay(and other) \\News with a Right Leaning View.
\#GaysGoRight\\
    \midrule
    \#GaysForTrump Support Our President \#GaysGoRight\\
    \midrule
This is the OFFICIAL Page of the \#GoRight Movement \\(due to loss of the previous pages we were forced to create a new page)\\
This is not the time to abandon the GOP it's time to \#GoRight and\\ create the Great American Revival of the GOP and America. \\
    \midrule
    \#GoRight and Join Us on \#TheQiew \\
When Gays Go Right We Make America Great \\
    \midrule
    Political organization https://gorightnews.com/\\
    \bottomrule
    \end{tabular}
    \caption{Current bios of top-5 coordinated pages in Facebook sorted by degree.}
    \label{tab:facebook-bios}
\end{table}

\begin{table}[t]
    \scriptsize
    \centering
    \setlength{\tabcolsep}{8pt}
    \renewcommand{\arraystretch}{1.2}
    \begin{tabular}{c}
    \toprule
    \textbf{Bio}  \\
    \midrule
    All of the articles and opinions shared here are not the opinions of HVU.\\ Use your own discernment.
All of HVU's links in one olace\\
https://linktr.ee/highvibesup
Family - Love - Learn\\
    \midrule
    All of the articles and opinions shared here are not the opinions of HVU. \\Use your own discernment.
All of HVU's links in one olace\\
https://linktr.ee/highvibesup
Family - Love - Learn\\
    \midrule
Karli Channel The Real Karli and X only KarluskaP is real\\ everyone else is being sued for using copyright avi artwork \\
    \midrule
    `' \\
    \midrule
Please post Tartaria architecture, flat earth\\
    \bottomrule
    \end{tabular}
    \caption{Current bios of top-5 cross-platform coordinated accounts sorted by degree.}
    \label{tab:coordinated-bios}
\end{table}

\begin{table*}[h!]
    \centering
    \scriptsize 
    \begin{tabular}{lp{10cm}}
        \toprule
        \textbf{Top-5 Messages} \\
        \midrule
        \parbox[t]{10cm}{ %Politics & 
            "US Secretary of State Antony Blinken Walking the Congressional Office Building Halls Surrounded by Police

            'War criminal', 'genocide secretary'
            
            Protestors reminding him of his complicity in crimes against humanity.
            
            RealWorldNewsChannel
            RealWorldNewsChat"} \\
        \addlinespace
        \parbox[t]{10cm}{ %News & 
            "Its sickening,  disgusting money laundering MF's. In a meantime the people of United States are suffering! Can't afford basic necessities!!"} \\
        \addlinespace
        \parbox[t]{10cm}{ % Jew & 
            "I like how zoinists shout poor me when they get caught out for there evil crimes 
            Normal Jews hate zoinist Jews 
            People of the world don't hate Jews 
            People of the world hate zoinists Jews n zoinists in general n what they stand for 
            They think there the chosen ones and that gives them the rite to kill n genecide inercent people 
            Zoinisium is hated by all races of the world 
            And has no place for peace in humanity "} \\
        \addlinespace
        \parbox[t]{10cm}{ % Ukraine & 
            "Ukrainians in Galway are being advised to vote for two Nigerians and a Labour candidate to best serve their interests. 

            'Under no circumstances vote for radicals - Irish Freedom Party, Independent Ireland, The Irish People and all others who have the slogans 'Ireland for the Irish'.
            
            \#ForeignInterference "} \\
        \addlinespace
        \parbox[t]{10cm}{ %Vaccine & 
        "Former CNN Anchor Chris Cuomo Admits to Suffering from a COVID Vaccine Injury
        
        ICYMI: There's been a major shift in the official narrative.

        Follow Vigilant\_News "} \\
        \addlinespace
            
        \addlinespace
        \bottomrule
    \end{tabular}
    \caption{Top-5 messages by total engagement for coordinated channels in Telegram}
    \label{tab:5-messages-telegram}
\end{table*}

\begin{table*}[h!]
    \centering
    \scriptsize 
    \begin{tabular}{lp{10cm}}
        \toprule
        \textbf{Top-5 Posts} \\
        \midrule
        \parbox[t]{10cm}{ % partisan
            "“WE THE PEOPLE” don’t play by your dictator rules. Just answer the damn question, crybaby. White House correspondents fire back after Biden snaps at reporter for refusing to 'play by the rules'"} \\
        \addlinespace
       \parbox[t]{10cm}{
            "If Biden wins there will no more arguing because the will not be anything left to argue about. He’s handing the keys over to illegals and the new world order. In the end, the joke will be on the democrats. The WHO/WEF hates them too and knows they’re imbeciles. Everyone knows"} \\
        \addlinespace
        \parbox[t]{10cm}{
            "'Shameful': GOP lawmaker shreds 'AWOL' Biden for throwing Jews 'under the bus' amid anti-Israel protests If this president, so-called president doesn’t personify evil destruction division of this country, I don’t know what or who would !"} \\
        \addlinespace
        \parbox[t]{10cm}{
            "Biden DHS docs suggested Trump supporters, military and religious people are likely violent terror threats. HaHaHa! I guess we also have Santa Claus/Easter bunny over here in training. Give me a break, we are trying to save this country from the bad guys!"} \\
        \addlinespace
        \parbox[t]{10cm}{
            "DeSantis spox dunks on NYT 'fact-check' on terrorists entering southern border: 'Awaiting your correction' They've been entering for last 3 years. Biden admin has no damn idea who is coming in. Even ones they think they know are using others' identity"} \\
        \addlinespace
        \bottomrule
    \end{tabular}
    \caption{Top-5 posts by total engagement for coordinated accounts in $\mathbb{X}$}
    \label{tab:5-messages-twitter}
\end{table*}

\begin{table*}[h!]
    \centering
    \scriptsize 
    \begin{tabular}{lp{10cm}}
        \toprule
        \textbf{Top-2 Posts} \\
        \midrule
        \parbox[t]{10cm}{
            "Youth Vote Shifts Toward Trump in 2024 Election https://gorightnews.com/youth-vote-shifts-toward-trump-in-2024-election/ Maybe the kids are alright... \#GoRightNews Recent polls indicate a surprising trend: young voters are warming to Donald Trump in the 2024 presidential election. The question arises: Are these voters aligning with Trump’s policies, or is President Biden driving them away? The answer might lie in a combination of both factors. Polling Data Reveals Shift According to the latest New York Times poll, young voters aged 18 to 29 favor Biden by a slim margin of two points, 47\% to 45\%. A Quinnipiac poll shows Trump leading Biden among voters aged 18 to 34, 48\% to 47\%. This is a stark contrast to the 2020 presidential election, where Joe Biden secured the youth vote by a significant 24\%. The last time a Republican won this demographic was in 1988. Biden’s Struggling Message Aidan Kohn-Murphy, founder of Gen Z for Change—a group that supported Biden in 2020—stated in the Washington Post, “Biden is out of step with young people on a number of key issues." Key issues where Biden seems to be losing support include: The War in Gaza: A majority of young voters, 51\%, support the Palestinians, while only 15\% support Israel. TikTok Ban: Biden's support for a TikTok ban is perceived as an attack on free speech. Consequently, 67\% of Gen Z voters say this makes them less likely to vote for him. Economic Challenges: High inflation and interest rates have made essential costs like food and housing unaffordable for young people entering the workforce or trying to purchase their first home. Trump’s Resonating Message The 45th president’s message appears to be resonating with young voters for several reasons: Gaza Conflict: While Trump supports Israel, he promises to bring a peaceful end to the conflict, citing his administration's four years of peace as evidence of his capability. TikTok Engagement: Trump recently joined TikTok, emphasizing that Biden wants to shut the platform down, thus appealing to younger users. Economic Performance: When asked about Trump’s handling of the economy, 65\% of young voters approved, compared to just 33\% for Biden. Social Media Dynamics According to CredoIQ, a social media analytics firm, nearly 25\% of the top left-leaning content creators on TikTok have posted anti-Biden content in the first four months of 2024, garnering over 100 million views. This content is often created by young, non-white liberals who share the belief that the U.S. Government, and specifically Joe Biden, aims to restrict free speech and information flow. Trump’s Adaptation to New Media In 2016, Trump broke political norms with an 'anyplace, anytime' approach, dominating cable TV. For the 2024 campaign, he has adapted this strategy to new media, appearing on podcasts, YouTube shows, and attending live events such as UFC and Formula 1. He has also made campaign stops at local venues, including bodegas, firehouses, and even in the South Bronx. The shift in the youth vote suggests a significant realignment in political affiliations, one that underscores the importance of addressing the issues most pertinent to young Americans. This analysis highlights the potential impact of these changing dynamics on the upcoming presidential election. [Source: Washington Post, Axios] https://archive.is/kk3HU https://www.axios.com/2024/06/13/trump-election-young-voters-polling \#GoRightNews Shared by Peter Boykin - American Political Commentator / Citizen Journalist / Activist / Constitutionalist for Liberty Web: https://PeterBoykin.com Kick: http://Kick.com/PeterBoykin YouTube: https://youtube.com/\@PeterBoykinForAmerica Twitter: https://twitter.com/GoRightNews Telegram: http://t.me/realpeterboykin Rumble: http://Rumble.com/GoRightNews Like the Content? Please Support! - Go Right News: Stripe: https://gorightnews.com/donations/support-gorightnews/ Cash App: http://cash.app/\$PeterBoykin1"} \\
        \addlinespace
        \midrule
        \parbox[t]{10cm}{
            "Rising Costs Highlight Challenges for American Families Cost of rent, energy, and other essentials surged in May In an alarming trend, the cost of essentials such as rent, energy, and groceries continues to surge, underscoring the persistent financial challenges faced by American families. While the overall Consumer Price Index (CPI) showed a slight stabilization, the specifics reveal a stark reality of escalating living expenses in our Constitutional Republic. Analyzing the Numbers The CPI indicated a 3.3\% rise in overall inflation compared to the previous year. Although this marks a slight decrease from April's 3.4\% and a significant drop from the 9.1\% peak in June 2022, it remains well above the Federal Reserve's target rate of 2\%. This persistent inflation underscores the ongoing economic strain on American households. Historical Context It's noteworthy that during the four years of Donald Trump's presidency, the average inflation rate was maintained at a modest 1.9\%. This comparison highlights a more stable economic period and suggests a need for policies that can effectively manage inflation without compromising the financial well-being of citizens. Essential Expenses on the Rise A closer examination of the May report reveals substantial increases in essential costs:  Rent: Up by 5.4\%  Mortgage: Up by 5.6\%  Hospital services: Up by 7.2\%  Car insurance: Up by 20.3\%  Electricity: Up by 5.9\%  Ground beef: Up by 4.9\%  Steak: Up by 5.7\%  Bacon: Up by 6.9\%  Hot dogs: Up by 7.3\% These increases in essential goods and services strain the budgets of American families, making everyday living increasingly unaffordable. Public Sentiment A recent poll reflects the public's discontent, with only 31\% of voters approving of President Biden's handling of inflation, while a significant 61\% disapprove. This sentiment underscores the urgent need for effective economic policies that address the real concerns of the populace. Administration's Response The Biden administration continues to assert progress in combating inflation. A statement from the White House on social media claimed: "Today’s report shows continued progress in lowering inflation. President Biden knows that costs are still too high for many families and we still have a lot more to do. That’s why he will keep fighting to lower drug costs, grocery prices, and energy bills." Critical Perspective However, many Americans find these assurances lacking. The everyday experience at grocery stores and gas stations starkly contrasts with the administration's optimistic declarations, leading to a disconnect between government rhetoric and public reality. As a Constitutional Republic dedicated to ensuring democracy and the well-being of its citizens, it is imperative that our government implements policies that stabilize the economy and reduce the financial burden on American families. The rising costs of essential goods and services are a pressing concern that requires immediate and effective action to safeguard the economic future of our nation. [Source: Poll, Whitehouse on X, BLS] https://d3nkl3psvxxpe9.cloudfront.net/documents/econTabReport\_maqVHQt.pdf https://x.com/WhiteHouse/status/1800957041390792843 https://www.bls.gov/news.release/pdf/cpi.pdf \#GoRightNews Shared by Peter Boykin - American Political Commentator / Citizen Journalist / Activist / Constitutionalist for Liberty Web: https://PeterBoykin.com Kick: http://Kick.com/PeterBoykin YouTube: https://youtube.com/\@PeterBoykinForAmerica Twitter: https://twitter.com/GoRightNews Telegram: http://t.me/realpeterboykin Rumble: http://Rumble.com/GoRightNews Like the Content? Please Support! - Go Right News: Stripe: https://gorightnews.com/donations/support-gorightnews/ Cash App: http://cash.app/\$PeterBoykin1"} \\
        \addlinespace
        \bottomrule
    \end{tabular}
    \caption{Top-2 posts by total engagement for coordinated pages in Facebook}
    \label{tab:5-messages-facebook}
\end{table*}

\begin{table*}[h!]
    \centering
    \scriptsize 
    \begin{tabular}{lp{10cm}}
        \toprule
        \textbf{Top-5 Messages} \\
        \midrule
        \parbox[t]{10cm}{ %MAGA & 
            "!You are a Golden Child!

            A Golden Child 94
            Unicorn 94 Third Eye 94 Covenant 94 Harmony 94
            Praise God 94 Divine Gene 94 Nine Six 94 Blue Eyes 94 Indigo Child 94 Ultra Maga 94 John John 94 White Hat 94 American Eagle 94 Carry On 94 New Earth 94 Pineapple 94
            
            PassionForFruit"} \\
        \addlinespace
        \parbox[t]{10cm}{ %Environment & 
            "\#CAWildfireSituationUpdate as of 6-16-2024, 5:03 P.M. PST.
            
            \#PostFire - 12,266 acres, 2\% containment
            \#HesperiaFire - 1,330, 7\%
            \#JunesFire - 1,076, 70\%
            \#JacksonFire - 876, ?\%
            \#HernandezFire - 600, 25\%
            \#MaxFire - 500, 0\%
            \#LisaFire - 350, 0\%
            \#PointFire 150, 0\%"} \\
        \addlinespace
        \parbox[t]{10cm}{ %Religion & 
            "What is it like to be wiser than your Creator? Job tried that and later repented in dust and ashes. Job was a wise and blessed man of faith."} \\
        \addlinespace
        \parbox[t]{10cm}{ % Conspiracy 
            "What was the true history behind all of these melted ruins? Certainly not the result of erosion, but perhaps a plasma storm? The results of a sudden flip in our polar magnetic electric field perhaps."} \\
        \addlinespace
        \parbox[t]{10cm}{ %Environment & 
            "WAVELAND, Mississippi 
        A boil advisory was issued Mon afternoon dt burst in the main water line.
        
        NASHVILLE, Indiana 
        A boil advisory was issued Mon after a water main break on Honeysuckle Ln. 
        
        MONTICELLO, Kentucky 
        A boil advisory was issued Mon for the East Hwy 92 area dt a water main break. 
        
        MILAN, Ohio
        Seminary Rd bw Perrin Rd/Broad St in Milan is closed dt a water main break. 
        
        SASKATOON, SK - Canada 
        North Park Wilson School is closed dt a water main break. 
        
        NEW CASTLE CO., Delaware 
        Shipley Rd at Foulk Rd was closed Mon night dt a water main break. 
        
        MONTGOMERY CO., Maryland 
        Dt a water main break Mon, a portion of the NW Branch Stream Valley Park near Highwood Terrace was undergoing repairs. 
        
        SIOUX FALLS, South Dakota 
        A water main break Mon afternoon caused flooding in downtown. 
        
        BOURBON CO., Kansas 
        There is a water main break at the Bourbon Co Transfer Station. 
        
        FORTUNA, California 
        Water will be shut off Wed to repair a broken water main on S Fortuna Blvd."} \\
        \addlinespace
        \bottomrule
    \end{tabular}
    \caption{Top-5 posts by total engagement for cross-platform coordinated accounts}
    \label{tab:5-messages-cross}
\end{table*}

\section{Topic Analysis}
We report the inferred topics and representative posts in each platform.

\begin{table*}[h!]
    \centering
    \scriptsize 
    \begin{tabular}{lp{10cm}}
        \toprule
        \textbf{Topic} & \textbf{Representative Tweets} \\
        \midrule
        rfk & \parbox[t]{10cm}{
            "If Kennedy iselected it will cost the evil doers over a trillion dollars in the first year. If Kennedy is in the debate it will show Biden as sinile, Trump as weak and Bobby will be elected. So they won't let Bobby in but will compensate CNN and anyone that might go to jail." \\
            "I’ve never been more convinced that both Trump and Biden fear Kennedy. The uniparty has no soul, and they hate democracy. We will never forget their attempts to silence us. Kennedy FTW"} \\
        \addlinespace
        debate & \parbox[t]{10cm}{
            "Biden’s really not up to debate is what this means. Whenever a question is asked, Biden’s promoters will either tell the answer in an earplug or use a prompter screen on the podium so Biden can read the answer. This is funny." \\
            "Trump has guts though. They tried to make it as unfriendly for him as possible, and he still is going to do it. I’m sure they were hoping he would say no way, and then Biden could brag about Trump being afraid to debate him."} \\
        \addlinespace
        covid-vaccine & \parbox[t]{10cm}{
            "Remember this on election day, remember All the "suddenly died" family and friends! Biden did this, he made vaccines mandatory! Just one of many bad Biden decisions!" \\
            "Trump was during Pandemic; FAILED to protect the American people when he knew damn well how infectious; deadly Covid Virus was, instead Trump told America "Covid was a Hoax"; failed to create a National Suppression Plan to save lives 700KAmericans DIED under Trump!"} \\
        \addlinespace
        bowman-latimer & \parbox[t]{10cm}{
            "Bowman called Biden a liar. Bowman voted against the Infrastructure Act. Bowman even voted against the Debt Limit increase that could have jeopardized Social Security. Bowman did not earn another term as a Democrat." \\
            "Bowman is his worst enemy. His and Rape denying has turned his district against him. His Votes AGAINST the Biden Administration's Progressive policies are NOT helping his district! Vote !"} \\
        \addlinespace
        border-security & \parbox[t]{10cm}{
            "with your lies. It is that BLOCKED Bipartisan Border Security Bill bc told GOP to "Kill The Bill". publicly stated it. The Border Crisis is NOW all on for Blocking Border Bill to SECURE THE BORDER!" \\
            "This bill was a sham and would not close border,but would allow more to take American jobs!! Donalds reiterates why GOP rejected ‘bipartisan’ border bill to head off potential debate talking point"} \\
        \addlinespace
        epstein-files & \parbox[t]{10cm}{
            "His flights with Epstein were with family present and years before he had an island. He's also the only one saying anything about releasing the logs, Biden sure isn't, and Trump took umbrage when DeSantis started advocating for their release for some strange reason..." \\
            " just another day in 'Murica, home of the child sex slaves. Trump made these folks feel they deserved Matt Gaetz' job, Trumps' job or Epstein's job if they just 'networked' enough. Investigate the real 'satanic' sex cult that police and politicians hide."} \\
        \addlinespace
        student-loan & \parbox[t]{10cm}{
            "This administration and president Biden seem to be answerable to nobody between giving money to student loan forgiveness even after the court ruled against them not getting Kennedy secret service protectiin, the list goes on no accountability." \\
            ""Summary: We estimate that President Biden\'s recently announced "New Plans" to provide relief to student borrowers will cost \$84 billion, in addition to the \$475 billion that we previously estimated for President Biden\'s SAVE plan.""} \\
        \addlinespace
        hunter-biden & \parbox[t]{10cm}{
            "Hunter Biden's laptop has nothing to do with the false Electoral College votes sent in for the 2020 election. Also, Alexander Smirnov (an FBI informant) was INDICTED for his false accusations about Barisma and the Bidens." \\
            "Actions speak louder than how they phrased something – especially when the actions are repeated over and over again. Their lies are proven in this week's Hunter Biden trial where the laptop is used as evidence." } \\
        \addlinespace
        blacks-against-biden & \parbox[t]{10cm}{
            "People aren't paying attention and don't expect him to be the President and don't know about Biden's accomplishments. That's what the polls show now but that's gonna change. Trump isn't gonna win 20\% of the black vote. No were here near it" \\
            "'At this stage in his presidency (3.5 years in) Biden must stop trying to convince Black voters his policies have made the US a significantly better place. Instead he should spend everything he’s got highlighting how anti-Black Trump is and how quickly any past gains will be lost."} \\
        \addlinespace
        \bottomrule
    \end{tabular}
    \caption{Topics and their representative posts on $\mathbb{X}$}
    \label{tab:topics_tweets}
\end{table*}

\begin{table*}[h!]
    \centering
    \scriptsize 
    \begin{tabular}{lp{15.5cm}}
        \toprule
        \textbf{Topic} & \textbf{Representative Messages} \\
        \midrule
        vaccine-skepticism & \parbox[t]{15.5cm}{
            "BREAKING: US doctors and scientists are currently investigating whether the COVID-19 virus is to blame for an “unusual pattern” of rare and deadly cancers that have been popping up in the wake of the pandemic.\@GeneralMCNews" \\
            "You, the unvaccinated are a special group of people. You stood your ground.  You did not let the pressures of society sway you.  You withstood the harshest discrimination seen in modern times.  You were excluded from society.  Some of you lost friendships and relationships with family members.  You are the best that society has to offer.  You used your analytical judgment and common sense.  You are the tree that withstood the hurricane.  You are superheroes! \@CovidVaccineTruth \@CovidVaccineTruthChat"} \\
        \addlinespace
        holistic-health & \parbox[t]{15.5cm}{
            "Ginger Lemonade for Immunity and Weight Loss\texttt{\textbackslash n}\texttt{\textbackslash n} We must:\texttt{\textbackslash n} - two large lemons\texttt{\textbackslash n} - A piece of ginger root (about 10-15 cm)\texttt{\textbackslash n} - two liters of chilled drinking water.\texttt{\textbackslash n} \texttt{\textbackslash n} Wash the lemon well and peel the ginger.  Cut the lemon and ginger into larger pieces and grind them in a blender.  Put everything in a jug, fill it with water and leave it in the fridge overnight." \\
            "HACKS: CAR SCRATCHES\texttt{\textbackslash n}\texttt{\textbackslash n}“A life hack you're gonna wish you knew sooner!!”\texttt{\textbackslash n}\texttt{\textbackslash n}I LOVE that our NATURAL REMEDIES are now in the category of MIND BLOWING, seriously who knew that VINEGAR \& COCONUT OIL could eliminate car scratches.\texttt{\textbackslash n}\texttt{\textbackslash n}Follow: \@DrBarbara\_ONeill"} \\
        \addlinespace
        presidential-debate & \parbox[t]{15.5cm}{
            "CNN is devastated after the debate. \texttt{\textbackslash n} "It\'s not just panic, it\'s pain."\texttt{\textbackslash n} hahahah.\texttt{\textbackslash n}\texttt{\textbackslash n} "I think there\'s a lot of people who are going to want to see him consider taking a different course now. There is time for this party to figure out a different way forward if he will allow us to do that. Um. But that was not what we needed from Joe Biden. it\'s personally painful for a lot of people. It\'s not just panic, it\'s pain. From what we saw tonight."\texttt{\textbackslash n}" \\
            "The Biden regime, and their lapdogs at CNN, are going extreme lengths to control all aspects of the debate.\texttt{\textbackslash n}\texttt{\textbackslash n} No outside media. No audience.\texttt{\textbackslash n}\texttt{\textbackslash n} They are trying to completely control public perception of the event.\texttt{\textbackslash n}\texttt{\textbackslash n}They want you to reject the evidence of your eyes and ears."} \\
        \addlinespace
        flat-earth-conspiracy & \parbox[t]{15.5cm}{
            "The Encyclopedia of Freemasonry reveals that the Earth is indeed flat. Organizations like The Freemasons know the truth about God and our flat earth but they choose to keep it hidden." \\
            "Flat Earth was the cosmology of our ancestors who had direct contact with the Divine because they were closer to the Divine than we are now. That is how the ancient's knew things despite not having access to technology we have now."} \\
        \addlinespace
        sinn-féin & \parbox[t]{15.5cm}{
            "Foreign 'police' mercenaries led by an ex-RUC thug against Irish Citizens?\texttt{\textbackslash n}\texttt{\textbackslash n}This is only going to end one way.\texttt{\textbackslash n}\texttt{\textbackslash n}Pack your bags Harris. The game is up.\texttt{\textbackslash n}\texttt{\textbackslash n}The Irish political class can pack their bags as well. This land is our land, it can never be sullied again by the treacherous dogs currently in office." \\
            "Just watch this \texttt{\textbackslash n}\texttt{\textbackslash n} COOLE COUNTY WEST MEATH BEING PLANTED AT 5 AM IN THE MORNING - LOOK AT THE MONEY PUMPED INTO THIS INVASION/ PLANTATION YOU WILL BEGIN TO UNDERSTAND THE FINANCIAL REWARD FOR SELLING YOUR SOUL BEING A TRAITOROUS, SELL OUT PIG \texttt{\textbackslash n}\texttt{\textbackslash n}JUDGEMENT DAY APPROACHES \texttt{\textbackslash n}\texttt{\textbackslash n} éiReGoBragh \texttt{\textbackslash n}\texttt{\textbackslash n} Fergus (Ferg) Power (@FergusPower1) \texttt{\textbackslash n}\texttt{\textbackslash n}These treacherous bastards won’t do this for the Irish people . Out children will have to work for years to afford a home yet LOOK AT THIS. My blood is boiling. \texttt{\textbackslash n}\texttt{\textbackslash n} Good to see that lad using the word invasion \& not using unvetted. We need to keep pressing to change the language around this. \texttt{\textbackslash n}\texttt{\textbackslash n} It is an invasion, we are being replaced, legal immigration is worse, we are at war."} \\
        \addlinespace
        trump-trial & \parbox[t]{15.5cm}{
            "If this is legit, it should wipe out Trump’s conviction. \texttt{\textbackslash n} Judge Juan Merchan has alerted Trump’s attorneys to a Facebook post by a supposed cousin of a Trump juror who spilled the beans that he had inside info that Trump was about to be convicted \texttt{\textbackslash n} "My cousin is a juror and says Trump is getting convicted. Thank you folks for all your hard work!”\texttt{\textbackslash n}\texttt{\textbackslash n} T edit: BY LAW, it would immediately be tossed. Remember those lessons in Constitutional Law our REAL POTUS/CiC promised us....\texttt{\textbackslash n}\texttt{\textbackslash n}" \\
            "The jurors were divided into 3 groups of 4. They found Trump gulity on 34 counts. I think we've entered the final stages in this sting operation."} \\
        \addlinespace
        bible-citations & \parbox[t]{15.5cm}{
            "Repent ye therefore, and be converted, that your sins may be blotted out, when the times of refreshing shall come from the presence of the Lord.\texttt{\textbackslash n} For more, please download\texttt{\textbackslash n}" \\
            "2 Chronicles 7:14 \texttt{\textbackslash n} \texttt{\textbackslash n} “If my people, which are called by my name, shall humble themselves, and pray, and seek my face, and turn from their wicked ways; then will I hear from heaven, and will forgive their sin, and will heal their land."} \\
        \addlinespace
        bird-flu-conspiracy & \parbox[t]{15.5cm}{
            "NIH-Funded Scientists Develop mRNA Bird Flu Vaccine ‘to Prevent Human Infections’ \texttt{\textbackslash n}\texttt{\textbackslash n} The likelihood of people getting H5N1 is very small, Dr. Robert Malone said. “The thing is, it doesn’t readily infect humans.”\texttt{\textbackslash n}\texttt{\textbackslash n}\texttt{\textbackslash n}" \\
            "Former coronavirus coordinator Deborah Birx wants to test millions of U.S. cows every week and screen dairy workers for "asymptomatic" cases of bird flu! \texttt{\textbackslash n}\texttt{\textbackslash n} You’ve got to be kidding me, this woman is nuts! They want to try and pull the same asymptomatic bs again with more fraudulent PCR tests! \texttt{\textbackslash n}\texttt{\textbackslash n}\texttt{\textbackslash n}" } \\
        \addlinespace
        ukraine-russia & \parbox[t]{15.5cm}{
            "JUST IN: Russia says it's ready to start arming enemies of the United States since the US is arming Ukraine.\texttt{\textbackslash n}\texttt{\textbackslash n} @BRICSNews" \\
            "They want to reassure Zelensky with the promise of a “bridge to the alliance” before the summit. The New York Times writes about this.\texttt{\textbackslash n}\texttt{\textbackslash n} NATO wants to give Kyiv “something significant” and at the same time maintain its position that it is not time for Ukraine to join the unification, the newspaper writes."} \\
        \addlinespace
        antisemitic-conspiracy & \parbox[t]{15.5cm}{
            "The left and right paradigm Israel is president is from Poland they're not real Jews most of them and even the masonic pedophile boys do Jewish rituals and then you got the Kabbalah which leads to Jewish mysticism black magic" \\
            "The Jews admit to controlling the world and they want to kill you and your children.\texttt{\textbackslash n}\texttt{\textbackslash n} It doesn't matter what you are, European, Arabian, African, Asian, Christian, Muslim or even atheist. No matter what you are, they want to kill you all. People must wake up and stop finghting with each other and realize who's their actual enemy!!!\texttt{\textbackslash n}\texttt{\textbackslash n} Follow us$>>$News\_Without\_Lies"} \\
        \addlinespace
        \bottomrule
    \end{tabular}
    \caption{Topics and their representative posts on Telegram}
    \label{tab:topics_messages}
\end{table*}

\begin{table*}[h!]
    \centering
    \scriptsize 
    \begin{tabular}{lp{12.5cm}}
        \toprule
        \textbf{Topic} & \textbf{Representative Posts} \\
        \midrule
        late-show\\against-trump & \parbox[t]{12.5cm}{
            "Stephen Colbert Speaks Truth to Trump's Most Damaging Lie - \#stephencolbert \#TheLateShow \#Trump \#DonaldTrump \#MAGA \#immigrants"\\
            "Stephen Colbert Speaks Truth to Trump's Most Damaging Lie - Rick Strom \#stephencolbert \#TheLateShow \#Trump \#DonaldTrump \#MAGA \#immigrants"} \\
        \addlinespace
        barron-trump-rnc & \parbox[t]{12.5cm}{
            "Former President Donald Trump said his youngest son Barron Trump was 17-years-old in an interview, when in fact he is 18 and a nominate delegate for the RNC." \\
            "Former President Donald Trump’s 18-year-old son, Barron Trump, is serving as a delegate to the Republican National Convention."} \\
        \addlinespace
        2020-election-acceptance & \parbox[t]{12.5cm}{
            "Republican National Committee co-chair Lara Trump said it's 'obvious' that Donald Trump has accepted the 2020 election results, when in fact, he has not." \\
            "Republican National Committee co-chair Lara Trump said it's 'obvious' that Donald Trump has accepted the 2020 election results, when in fact, he has not."} \\
        \addlinespace
        trump-trial-dayoff & \parbox[t]{12.5cm}{
            "Judge Juan Merchan gave Donald Trump the day off from his hush money trial to attend Barron Trump's graduation ceremony, but evidence suggests that he is attending political fundraiser." \\
            "Judge Juan Merchan gave Donald Trump the day off from his hush money trial to attend Barron Trump's graduation ceremony, but evidence suggests that he is attending political fundraiser."} \\
        \addlinespace
        jacked-up-biden & \parbox[t]{12.5cm}{
            "Rep. Greg Murphy, a Republican, says he has evidence that President Joe Biden was using "something" before his popular State of the Union address." \\
            "Rep. Greg Murphy, a Republican, says he has evidence that President Joe Biden was using "something" before his popular State of the Union address."} \\
        \addlinespace
        trump-classified-documents & \parbox[t]{12.5cm}{
            "More bombshell evidence dropped against former President Donald Trump in his classified documents case, just as Judge Aileen Cannon delayed the trial indefinitely." \\
            "More bombshell evidence dropped against former President Donald Trump in his classified documents case, just as Judge Aileen Cannon delayed the trial indefinitely."} \\
        \addlinespace
        rfk & \parbox[t]{12.5cm}{
            "Exciting Announcement! Join us for a special interview with independent presidential candidate RFK Jr. on Monday, June 17, immediately following The Young Turks! Tune in on tyt.com/live YouTube, Facebook, or Twitch! TYT members get exclusive access to a members-only Bonus Episode after the interview, where Cenk will ask even more personal questions and dive deeper into the issues! Sign up as a member today at  to join the discussion! Don't miss this unique opportunity to engage with crucial political dialogue and watch TYT hold a presidential candidate accountable! See you there!" \\
            "Exciting Announcement! Join us for a special interview with independent presidential candidate RFK Jr. on Monday, June 17, immediately following The Young Turks! Tune in on tyt.com/live YouTube, Facebook, or Twitch! TYT members get exclusive access to a members-only Bonus Episode after the interview, where Cenk will ask even more personal questions and dive deeper into the issues! Sign up as a member today at  to join the discussion! Don't miss this unique opportunity to engage with crucial political dialogue and watch TYT hold a presidential candidate accountable! See you there!"} \\
        \addlinespace
        coulter-ramaswamy & \parbox[t]{12.5cm}{
            "Conservative Ann Coulter told former Republican presidential candidate Vivek Ramaswamy said while she agrees with him on many issues, she wouldn't vote for him because he’s “an Indian.”" \\
            "Conservative Ann Coulter told former Republican presidential candidate Vivek Ramaswamy said while she agrees with him on many issues, she wouldn't vote for him because he’s “an Indian.”" } \\
        \addlinespace
        trump-falsifying-records & \parbox[t]{12.5cm}{
            "Fox News' Jeanine Pirro and conservative commentator Charlie Kirk are melting down over former President Donald Trump's 34-count conviction for falsifying business records in his hush money trial."} \\
        \addlinespace
        bannon-sentence & \parbox[t]{12.5cm}{
            "Donald Trump adviser Steve Bannon proclaimed that \'"no one" will shut him up as he was ordered to report to prison for refusing to testify before congress."} \\
        \addlinespace
        \bottomrule
    \end{tabular}
    \caption{Topics and their representative posts on Facebook}
    \label{tab:topics_posts}
\end{table*}

%\subsection{Facebook}
\section{Characterization of AI Generated Content}
Regarding the AI-Generated Content (AIGC) results, we conducted a qualitative analysis to characterize AIGC across platforms. 
Specifically, we analyzed the top 20 AIGC posts by popularity from each platform and used GPT-4 with zero-shot learning to estimate a partisan score. 
Our findings show that Telegram has an average partisan score close to zero, while $\mathbb{X}$ and Facebook exhibit higher scores of 0.5 and 0.4, respectively. This highlights differences among AIGC diffusion between the three platforms that help to interpret and characterize the results.

\end{document}